\documentclass[pdflatex,sn-mathphys-num]{sn-jnl}

\usepackage{graphicx}%
\usepackage{multirow}%
\usepackage{amsmath,amssymb,amsfonts}%
\usepackage{amsthm}%
\usepackage{mathrsfs}%
\usepackage[title]{appendix}%
\usepackage{xcolor}%
\usepackage{textcomp}%
\usepackage{manyfoot}%
\usepackage{booktabs}%
\usepackage{algorithm}%
\usepackage{algorithmicx}%
\usepackage{algpseudocode}%
\usepackage{listings}%
\usepackage{tikz}

\theoremstyle{thmstyleone}%

\theoremstyle{thmstyletwo}

\theoremstyle{thmstylethree}

\newcommand{\R}{\mathbb{R}}
\newcommand{\N}{\mathbb{N}}
\newcommand{\Z}{\mathbb{Z}}
\newcommand{\C}{\mathbb{C}}
\newcommand{\lset}[1]{\left\{ #1 \right\}}
\newcommand{\btheta}{\boldsymbol \theta}
\newcommand{\bOmega}{\boldsymbol \Omega}
\newcommand{\bx}{\boldsymbol x}

\begin{document}

\title[QNN Benchmarking]{Benchmarking Encoding Families in Quantum Neural Networks Under Fixed Circuit Area for Frequency Spectrum and Trainability}

\author*[1,2]{\fnm{Martyna} \sur{Czuba}}\email{martyna.czuba@wat.edu.pl}

\author[3]{\fnm{Patrick} \sur{Holzer}}\email{patrick.holzer@itwm.fraunhofer.de}

\author[2,4]{\fnm{Hein Zay Yar} \sur{Oo}}\email{heinzayyar.o@thewiser.org}

\affil*[1]{\orgname{Military University of Technology},
  \orgaddress{\street{gen. Sylwestra Kaliskiego 2},\postcode{00-908} \city{Warsaw}, \country{Poland}}}

\affil[2]{\orgname{The Washington Institute for STEM Entrepreneurship and Research},
  \orgaddress{\city{Washington, DC}, \country{USA}}}

\affil[3]{\orgdiv{Department for Financial Mathematics},
  \orgname{Fraunhofer Institute for Industrial Mathematics ITWM},
  \orgaddress{\street{Fraunhofer-Platz 1}, \city{Kaiserslautern}, 
              \postcode{67663}, \country{Germany}}}
              
\affil[4]{\orgdiv{Department of Computer Science},\orgname{ Georgia Institute of Technology},\orgaddress{\street{225 North Ave NW},\city{Atlanta}, \state{Georgia}, \postcode{30332}, \country{United States of America}}}

\abstract{  
Quantum Neural Networks (QNNs) 
offer a promising framework for integrating quantum computing principles into machine learning, yet their practical capabilities and limitations remain insufficiently studied. 
In this work, we systematically investigate the trainability and approximation properties of QNNs by benchmarking diverse circuit architectures and encoding strategies across synthetic and real-world datasets.
We analyze several ansätze, including Hamming, binary, exponential, ternary, turnpike and Golomb, by evaluating their ability to learn synthetic data modeled as random finite Fourier series. To assess real-world applicability, we further evaluate QNNs on two time-series classification tasks: a Fischertechnik pneumatic leak detection dataset and the publicly available NASA bearing fault dataset.
Our experiments show that while broader frequency spectra can theoretically enhance expressivity, practical trainability is strongly influenced by architectural factors such as qubit count and circuit depth. Notably, we find that QNNs perform best when the frequency spectrum is tailored to the target function’s complexity but remains as compact as possible. Moreover, architectures with identical frequency spectra can differ in trainability, with configurations using more qubits and fewer layers generally performing better, except in the single-layer case.
These findings provide guidelines for selecting QNN ansätze and offer new insights into the interplay between expressivity and trainability in quantum machine learning.}




\maketitle
\section{Introduction}\label{sec:introduction}

Quantum Machine Learning (QML) combines Quantum Computing (QC) principles with classical Machine Learning (ML) techniques to take advantage of quantum mechanics. It is still an open question whether QML models offer an advantage over classical models, especially since there are theoretical and practical hurdles in comparison, as real-world problems beyond cryptography and quantum chemistry are often mathematically not rigorously formalizable and many practical problems can already be solved today with classical hardware and algorithms \cite{schuld2022quantumadvantage}. Nevertheless, hybrid QML algorithms in particular, which are currently considered the most promising candidates to run on the currently available Noisy Intermediate-Scale Quantum (NISQ) hardware, offer the opportunity to gain new insights into ML, to combine the advantages of classical and quantum algorithms and to develop new quantum-inspired classical algorithms \cite{schuld2022quantumadvantage, callison2022hybrid, wolf2024careaboutquantummachine, pulicharla2023hybrid, Preskill2018QuantumCI}.
At its core, QML often uses parameterized quantum circuits (PQCs) - such as quantum neural networks (QNNs) \cite{holzer2024specinv, Schuld2020EffectOD, Farhi2018ClassificationWQ, McClean2018BarrenPI}, variational quantum algorithms (VQAs) \cite{Cerezo2020VariationalQA, Tilly2021TheVQ}, Quantum Alternating Operator Ansatz (QAOA) \cite{Hadfield2017FromTQ} and Variational Quantum Eigensolver (VQE) \cite{Tilly2021TheVQ, Kandala2017HardwareefficientVQ} - to encode data into high-dimensional quantum states and extract information through optimised unitary transformations. These models are particularly suitable for NISQ devices, where hybrid algorithms allow quantum circuits to process information while classical systems perform parameter optimization. 

In this work, we focus on the trainability of QNNs, which we experimentally evaluate using synthetic data and two real data sets. QNNs are parameterized circuits, encoding the training data $\textbf{x} = (x_1,...,x_N) \in \R^{N}$ and the trainable parameters $\mathbf{\theta}=(\theta_1,...,\theta_P) \in \R^P$. More precisely, the circuit of a QNN consists of alternating layers of data encoding circuits $S_l(\textbf{x})$ and parameter encoding circuits $W_{\mathbf{\theta}}^{(l)}$, where the data encoding circuits typically have the form $S_l(\textbf{x}) = e^{-ixH_l}$ for some Hamiltonian $H_l$ called the \textit{generator}. Furthermore, it is often assumed in the literature that $H_l$ itself is composed of smaller Hamiltonians, which \cite{holzer2024specinv} has called \textit{sub-generators} in order to differentiate them from the generators \cite{Schuld2020EffectOD, Kordzanganeh2022AnEF, Shin2022ExponentialDE}.
It can be shown that the repeated use of data embeddings, so-called re-uploading or input redundancy, is necessary to increase the expressivity of the model \cite{Kandala2017HardwareefficientVQ, Vidal2019InputRF, PerezSalinas2019DataRF}. More details about the architecture and properties of QNNs are described in Section~\ref{sec:qnn-general}.

The question arises which functions can be represented or approximated by QNNs. 
It can be shown that QNNs are equivalent to classical finite Fourier series \cite{holzer2024specinv, Schuld2020EffectOD}. Consequently, any function with sufficient properties can be approximated by QNNs with arbitrary precision, provided that the observable and the parameter encoding layers $W_{\mathbf{\theta}}^{(l)}$ can be chosen arbitrarily and adapted to the target function \cite{Schuld2020EffectOD}. While approximability is therefore given in theory, the question of approximation and training speed as well as approximation quality arises in practice. As in the architecture of a neural network in classical machine learning, the observable is typically defined before training and is therefore not trained itself. Rather, the question is whether any function can be approximated under a given fixed observable or classes of observables, as questioned in \cite{holzer2024specinv}. Furthermore, the parameter embedding circuits are not arbitrarily selectable or trainable in practice but typically consist of a few entangling layers and a few trainable parameters, so that only a small subspace of unitary operators is actually achievable. Thirdly, the so-called \textit{frequency spectrum}, i.e. the set of all frequencies in the finite Fourier series, depends on the selected generators or sub-generators. On the one hand, you want the largest possible frequency spectrum as functions can be better approximated with larger Fourier series, provided that the coefficients of the Fourier series can be arbitrarily chosen. On the other hand, the coefficients cannot be arbitrarily chosen or trained for the reasons mentioned above, and are even further restricted by a lower doubling of the frequencies, so-called \textit{degeneracy}, so that it is unclear whether a rather large or a rather small frequency spectrum is advantageous in practice. In this work, we therefore want to investigate this question experimentally and analyse the approximation properties of various QNN ansätze using both synthetic data and real data.

\subsection{Related Work}
\label{sec:related-work}
The theory of the frequency spectrum of QNNs has been studied, among others, in \cite{holzer2024specinv, Schuld2020EffectOD,Heimann2022LearningCO, Peters2022GeneralizationDO,Kordzanganeh2022AnEF, Shin2022ExponentialDE}. We mainly follow the work of \cite{holzer2024specinv}, where different ansätze to QNNs, namely different families of sub-generators, were studied. We summarize the theoretical results to provide a basis for understanding the experimental setup in Section~\ref{sec:qnn-general}.

For benchmarking different ansätze, we make use of the \emph{learning capability} metric introduced in \cite{Heimann2022LearningCO}. The learning capability \( \mu_K \) is defined as:
\begin{equation}
\mu_K := \frac{1}{|G_K|} \sum_{g \in G_K} \varepsilon_g,
\end{equation}
where \( G_K \) represents a set of \( |G_K| \) many randomly chosen and normalized Fourier series with fixed finite frequency spectrum \( K \), and \( \varepsilon_g \) denotes the individual loss associated with each function. More precisely, we utilize the mean squared error (MSE) as the primary loss metric, defined as:
\begin{equation}
\varepsilon_g = \frac{1}{|X|} \sum_{x \in X} \left( f_{\boldsymbol{\theta}}(x) - g(x) \right)^2,
\end{equation}
where \( |X| \) is the size of the dataset, \( f_{\boldsymbol{\theta}}(x) \) is the output of the QNN model parameterized by the trained \( \theta \), and \( g(x) \) is the target Fourier series. Due to the Fourier nature of QNNs, this metric is a sufficient choice to benchmark the learning capabilities.

\subsection{Structure of the Paper}
The remainder of this paper is organized as follows. Section~\ref{sec:qnn-general} repeats the foundations of Quantum Neural Networks (QNNs), explaining their architecture, the theoretical framework of frequency spectra, and various encoding strategies. Section~\ref{sec:learning-capability} evaluates the learning capability of QNNs using synthetic data. Section~\ref{sec:fischertechnik-dataset} presents experiments on the Fischertechnik Fabrik Dataset, a real-world industrial application scenario for pneumatic leak detection. Section~\ref{sec:nasa-bearing-dataset} extends our evaluation to the NASA bearing dataset. Finally, we summarize our findings, discuss practical implications, and outline directions for future research in Section~\ref{sec:conclusion}.

\section{Quantum Neural Networks}
\label{sec:qnn-general}
Quantum Neural Networks are defined as the expectation value of some specific parameterized quantum circuit
\begin{equation}
    f_{\btheta}(\bx) = \langle 0| U^\dagger(\bx,\btheta) M U(\bx,\btheta)|0\rangle,
\end{equation}
where $M$ is a fixed observable. Typically, $M$ is selected as a Pauli string, often even simply as $P \otimes \text{Id} \otimes \cdots \otimes \text{Id}$, where $P \in \{X, Y, Z\}$ denotes some Pauli operator. In this work and in all experiments, we fix $M=Z\otimes \text{Id} \otimes \cdots \otimes \text{Id}$, hence measuring only the first qubit.
The parameterized unitary $U(\bx,\btheta)$ consists of multiple layers, each containing a data encoding layer and a parameter encoding layer. For \textit{univariate functions}, i.e. $x \in \R$, the circuit is of the form
\begin{equation}
     U(x,\btheta) = W^{(L+1)}_{\btheta}\underbrace{S_L(x)W^{(L)}_{\btheta}}_{\text{Layer } L}\cdots W^{(2)}_{\btheta} \underbrace{S_1(x)W^{(1)}_{\btheta}}_{\text{Layer } 1},
\end{equation}
where $R \in \N$ denotes the number of qubits, $L\in \N$ the number of layers, $W^{(l)}_{\btheta}$ the parameter encoding layers and $S_l(x)$ the data encoding layers for $l=1,...,L$.
Before we take a closer look at the data encoding layers $S_l(x)$, we must first clarify how multivariate models with $\bx\in \R^N$ for $N>1$ can be built from univariate models. There are several approaches, two of which are called \emph{sequential} and \emph{parallel ansatz}. In the sequential ansatz we have circuits $U(x_n,\btheta)$ in the previously mentioned layered form for each variable $x_n \in \R$ and execute them sequentially. Specifically, 
\begin{align}
    U(\bx, \btheta) := \prod_{n=1}^N U_n(x_n, \btheta).
\end{align}
In this case, the circuit still consists of $R$ qubits but $N \cdot L$ many layers. Analogously, in the parallel ansatz the circuits $U_n(x_n, \btheta)$ are executed in parallel, which is realized in quantum mechanics by the tensor product. However, in the parallel ansatz we only glue the data encoding layers $S_{l, n}(x_n)$ together, i.e. 
\begin{align}
\label{eq: parallel ansatz}
     S_l(\bx) := \bigotimes_{n=1}^N S_{l, n}(x_n).
\end{align}
In this case, the circuit consists of $ N \cdot R$ many qubits and $L$ many layers.

The data encoding layers have the general form
\begin{align}
    S_l(x)=e^{-ix H_l}
\end{align} 
for some Hamiltonians $H_l$ called \emph{generators}. Typically, the $H_l$ are chosen such that they are made up of operators acting only on a few or even a single qubit, which are consequently called \textit{sub-generators}. For more details, see \cite{holzer2024specinv}.

It can be shown that QNNs can be represented as finite Fourier series
\begin{equation}
    f_{\boldsymbol{\theta}}(\boldsymbol{x}) = \sum_{\boldsymbol{\omega} \in \bOmega} c_{\boldsymbol{\omega}}(\boldsymbol{\theta}) e^{i \boldsymbol{\omega} \cdot \boldsymbol{x}},
    \label{eq:fourier_series}
\end{equation}
where \( \boldsymbol{\omega} \cdot \boldsymbol{x} \in \mathbb{R} \) denotes the standard scalar product between \( \boldsymbol{\omega} \) and \( \boldsymbol{x} \), and \( \bOmega \subseteq \mathbb{R}^N \) is a finite set referred to as the \emph{frequency spectrum} \cite{holzer2024specinv, Schuld2020EffectOD}. The frequency spectrum $\bOmega$ depends only on the generators or sub-generators used and is independent of whether the parallel or sequential ansatz was used, while the coefficients $c_{\boldsymbol{\omega}}(\boldsymbol{\theta})$ also depend on the parameter embedding circuits and the observable $M$. For a univariate model, the frequency spectrum can be expressed very compactly algebraically as follows.
Let 
\begin{equation}
    \Delta S := \lset{s_1-s_2|\ s_1,s_2 \in S}
\end{equation}
and 
\begin{equation}
\sum_{l=1}^L S_l := \lset{ \sum_{l=1}^L \lambda_l \mid (\lambda_1,\ldots,\lambda_L)\in S_1\times\cdots\times S_L }
\end{equation}
for all sets $S, S_1,...,S_L\subseteq \R^N$, and let $\sigma\left(H_{l}\right) \subseteq \R$ be the spectrum of $H_l$, then the frequency spectrum of the univariate model is given by
\begin{align}
    \Omega  = \Delta \sum_{l=1}^L \sigma\left(H_l\right).
\end{align}
If the generator $H_l$ is made up of sub-generators (in a meaningful way), the sum is taken over all sub-generators. The frequency spectrum of a multivariate QNN is the Cartesian product  $\bOmega = \Omega_1\times \cdots \times \Omega_N$ of the frequency spectra $\Omega_n = \Delta\sum_{l=1}^L \sigma\left(H_{n, l}\right) $ of the associated univariate models. 

The algebraic structure of the frequency spectrum has interesting implications. If we define an invariant called the \textit{area} $A$ of the univariate QNN as $A:=R \cdot L$, and the \textit{shape} as $(R, L)$, then all QNNs with the same (sub-)generators with the same area will have the same frequency spectrum. This means that the specific arrangement, in particular the shape, is freely choosable as long as the area is invariant, at least as far as the frequency spectrum is concerned. This property is called \textit{spectral invariance under area preserving transformations} \cite{holzer2024specinv}. 

However, the size of the frequency spectrum is only one factor influencing the approximation properties of the QNN. Although finite Fourier series can approximate functions increasingly better as the number of frequencies increases, as long as the coefficients can be chosen completely freely, the latter condition is not the case with QNNs. As already described, the coefficients depend on the frequency doublings, the so-called degeneracy, and the parameter encoding circuits. If there is a high degeneracy, the frequency spectrum is smaller but there is more freedom in the coefficients, and conversely, a large frequency spectrum leads to less freedom in the coefficients. Although the signature can be chosen freely with respect to the frequency spectrum for a given area, the approximation quality of QNNs under different signatures with a constant area will be analyzed in the following. Different ansätze that lead to frequency spectra of different sizes will also be analyzed. In \cite{holzer2024specinv, Schuld2020EffectOD, Heimann2022LearningCO}, the so called \textit{Hamming encoding} was studied, in which all sub-generators were selected as constant $\tfrac{Z}{2}$. Note that $Z$ can be replaced by any other Pauli matrix, but we fix $Z$ through all of our experiments. The univariate frequency spectrum using Hamming encoding is given by
\begin{equation}
   \Omega =  \Z_{A},
\end{equation}
where $A=R \cdot L$ again denotes the area of the QNN and $\Z_K:= \{-K,...,0,...,K\}$ for all $K \in \N_0$. The size of the frequency spectrum is therefore linear in the area $A$. Further ansätze with operators that act on individual qubits are obtained by additionally allowing different prefactors before $\tfrac{Z}{2}$, that is, we consider $H_{r, l}$ of the form $H_{r,l} = \beta_{r, l} \cdot \tfrac{Z}{2}$ with $\beta_{r,l} \in \R$. Only the scaling of the Hamiltonians varies, while the circuit complexity remains unchanged.
If $\beta_{r, l} := 2^{(l-1) + L \cdot (r-1)}$ is chosen as consecutive powers of 2, the encoding is called \textit{binary encoding}, and the resulting QNN has the frequency spectrum
\begin{align}
    \Omega = \Z_{2^A-1},
\end{align}
and its size is therefore exponential in the area $A$ \cite{holzer2024specinv, Peters2022GeneralizationDO}. Modifying only the largest factor in this encoding to $\beta_{R, L} := 2^{A -1} + 1$ if $A >1$, which is then called \textit{exponential encoding}, the frequency spectrum is now slightly larger with 
\begin{equation}
     \Omega = \Z_{2^A},
\end{equation}
see \cite{holzer2024specinv, Kordzanganeh2022AnEF}. 
The question is, what is the maximum frequency spectrum one can achieve? It can be shown \cite{holzer2024specinv, Shin2022ExponentialDE, Peters2022GeneralizationDO} that under the assumption that all generators act on single qubits, the choice of $\beta_{r, l} := 3^{(l-1) + L \cdot (r-1)}$, called the \textit{ternary encoding}, leads to a maximal frequency spectrum of
\begin{equation}
    \Omega = \Z_{\frac{3^A - 1}{2}},
\end{equation}
where the frequency spectrum is maximal in two senses:
\begin{enumerate}
    \item There is no other QNN with frequency spectrum $\Omega'$ such that $\Z_K \subseteq \Omega'$ and $K > \frac{3^A-1}{2}$ (called \textit{maximality in $K$}).
    \item There is no other QNN with frequency spectrum $\Omega'$ such that $|\Omega'| > |\Omega| = 3^A$ (called \textit{maximality in size}).
\end{enumerate}

While the first definition asks for the largest possible gap-free frequency spectrum around 0 included in $\Omega$, the second is concerned with the largest frequency spectrum in terms of quantity. Under the condition that the sub-generators may only act on individual qubits, the two terms therefore coincide \cite{holzer2024specinv}. However, if one allows the sub-generators to operate on several qubits, these terms of maximality no longer coincide. Two ansätze can be mentioned here. Let $H_l$ be decomposed into subgenerators $H_{1,l}, \ldots, H_{\tfrac{R}{q},l}$, each acting on $q$ qubits. 
Furthermore, set $H_{r, l} = \beta_{r,l} \cdot H$,where $H$ is a fixed Hamiltonian. In the \textit{turnpike encoding} we define
\begin{equation}
    \beta_{r,l} := (2K+1)^{(l-1) + L \cdot (r-1)},
\end{equation}
where $K:=\max \{K' \in \N_0|\ \Z_{K'} \subseteq \Delta\sigma(H)\}$ and $H$ is defined as a Hermitian matrix with some eigenvalues $\lambda_1,...,\lambda_{2^q} \in \Z$. Then
\begin{equation}
    \Z_{\frac{(2K+1)^{\frac{A}{q}}-1}{2}} \subseteq \Omega
\end{equation}
holds and the frequency spectrum is $K$-maximal, if $L=1$, $q = R$ and the eigenvalues 
$\lambda_1,...,\lambda_{2^q} \in \Z$ of $H$ are a solution of the so called relaxed turnpike problem introduced in \cite{holzer2024specinv}. If $L>1$, $q <R$ or the eigenvalues are not a solution to the relaxed turnpike problem, the frequency spectrum is not necessarily $K$-maximal.

In the \textit{Golomb encoding}, one again fixes some Hermitian matrix $H$ with some eigenvalues $\lambda_1\leq ...\leq\lambda_{2^q} \in \Z$ and defines
\begin{equation}
    \beta_{r,l} := (2\ell(\sigma(H))+1)^{(l-1) + L \cdot (r-1)},
\end{equation}
where $\ell(\sigma(H)) := \lambda_{2^q} - \lambda_1$. In this case, the resulting frequency spectrum has size
\begin{equation}
    |\Omega| = |\Delta \sigma(H)|^{\tfrac{A}{q}},
\end{equation}
which is maximal in size if and only if the eigenvalues of $H$ are a so called Golomb ruler \cite{holzer2024specinv, Peters2022GeneralizationDO, Kordzanganeh2022AnEF}, in that case $|\Delta \sigma(H)| = (4^q-2^q+1)$.

In this work, we study the practical approximation properties of all the previously mentioned ansätze on different datasets.

\section{Learning Capability}
\label{sec:learning-capability}
\subsection{Dataset}

As mentioned previously, the output of a QNN can be expressed as the
finite Fourier series \( f_{\boldsymbol{\theta}}(\boldsymbol{x}) =
\sum_{\boldsymbol{\omega} \in \Omega}
c_{\boldsymbol{\omega}}(\boldsymbol{\theta}) e^{i \boldsymbol{\omega} \cdot \boldsymbol{x}} \),
see Eq.~\eqref{eq:fourier_series}.
To evaluate the QNNs performance,we adopt the \textit{learning capability} metric introduced in \cite{Heimann2022LearningCO}. This metric quantifies the model's ability to approximate univariate Fourier series with randomly selected coefficients and is defined as
\begin{equation}
   \mu_K := \frac{1}{|G_K|} \sum_{g \in G_K} \varepsilon_g, 
\end{equation}
where \( G_K \) is a set of normalized Fourier functions with frequency spectrum $\Z_K$, and 
\begin{equation}
\varepsilon_g := \frac{1}{|X|} \sum_{x \in X} \left( f_{\boldsymbol{\theta}}(x) - g(x) \right)^2,
\end{equation}
is the individual mean squared error loss associated with the function
\begin{align}
    g(x)=\sum_{k=-K}^{K} c_k e^{ikx}.
\end{align}
Note that a smaller learning capability means a better approximation property of the QNN.
For our experiments, we test four different frequency spectra with $K=2^2, 2^4, 2^6, 2^8$.   
For each function $g \in G_K$, the real and imaginary part of the Fourier coefficient $c_k \in \C$ was sampled individually from a normal distribution with mean 0 and variance 1 for all $k = 1,...,K$, while the constant term $c_0$ was drawn from a uniform distribution on the interval $(-0.7, 0.7)$ to ensure baseline variability. We generated $100$ univariate random Fourier series in this way. 

For each $g \in  G_K$, we divide the interval $[0, 2 \pi]$ into 4000 equidistant points and collect them in the dataset $X$ and define the target as $y := g(X)$.
Since we have fixed the observable $M=Z \otimes \text{Id} \otimes ... \otimes \text{Id}$, the QNN can only take values $f_{\boldsymbol{\theta}}(x) \in [-1, 1]$ because the eigenvalues of $M$ are all $\pm 1$, we have to scale $y$ eventually. For that, we always apply Min-Max normalization to scale $y$ in the range of $[-\frac{1}{2}, \frac{1}{2}]$. We have chosen the boundaries $\pm \frac{1}{2}$ in our experiments because it is a difficult problem to train Fourier series $g$, which are frequently close to the theoretical boundaries $\pm 1$.

\subsection{Model}
\label{subsec:learning-capability-model}
We computed the learning capability of all mentioned ansätze, that is, Hamming, binary, exponential, ternary, turnpike and Golomb. If $R \equiv 0 \mod 3$, we have chosen $q=3$ and $H:= \text{diag}(0, 8, 15, 17, 20, 21, 31, 39)$ for the turnpike ansatz since its eigenvalues are a solution to the relaxed Turnpike problem of size $8 = 2^3$ with $K=24$ \cite{holzer2024specinv}, and $H:= \text{diag}(0, 1, 4, 9, 15, 22, 32, 34)$ for the Golomb ansatz, as its eigenvalues are a Golomb ruler of size $8$ \cite{Babcock1953IntermodulationII}. Otherwise, if $R \equiv 0 \mod 2$, we set $q = 2$ and $H= \text{diag}(0,1,4,6)$, which is both a solution of the relaxed Turnpike problem and a perfect Golomb ruler \cite{holzer2024specinv, Babcock1953IntermodulationII}. In this case, the Golomb and Turnpike ansatz lead to the same QNN.

Additionally, different combinations of the shape $(R, L)$ with areas $A=2, 4, 6$ were examined, specifically 
\[(R, L) =(1,2), (2,1), (1,4), (2,2),(4,1), (1, 6), (2, 3), (3, 2), (6, 1). \]

The number of positive frequencies $|\Omega_{>0}|$ for each ansatz and shape can be found in Table~\ref{tab:frequency-spectra}.
\begin{table}[ht]
\caption{The size of the positive frequency spectrum for each ansatz and shape considered}
\label{tab:frequency-spectra}
\begin{tabular*}{\textwidth}{@{\extracolsep\fill}llrrrrrr}
\toprule
\multirow{2}{*}{Area} & \multirow{2}{*}{Shape} &
  \multicolumn{6}{c}{$|\Omega_{>0}|$} \\
\cmidrule(lr){3-8}
& & Hamming & Exponential & Binary & Ternary & Golomb\footnotemark[1] & Turnpike\footnotemark[1] \\
\midrule
2 & (1, 2) & 2 & 4 & 3 & 4 & 6 & 6 \\
  & (2, 1) & 2 & 4 & 3 & 4 & 6 & 6 \\
\midrule
4 & (1, 4) & 4 & 16 & 15 & 40 & 84 & 84 \\
  & (2, 2) & 4 & 16 & 15 & 40 & 84 & 84 \\
  & (4, 1) & 4 & 16 & 15 & 40 & 84 & 84 \\
\midrule
6 & (1, 6) & 6 & 64 & 63 & 364 & 1098 & 1098 \\
  & (2, 3) & 6 & 64 & 63 & 364 & 1098 & 1098 \\
  & (3, 2) & 6 & 64 & 63 & 364 & 1624 & $\ge 1200$ \\
  & (6, 1) & 6 & 64 & 63 & 364 & 1624 & $\ge 1200$ \\
\bottomrule
\end{tabular*}
\footnotetext[1]{The Golomb and Turnpike ansatz was only computed for $R\equiv 0 \mod 3$ and $R\equiv 0 \mod 2$, since $R$ must be divisible by $q$, which we set to $q=2$ or $q=3$ depending on $R$.}
\end{table}

\begin{figure}[t]
    \centering \includegraphics[width=0.98\textwidth]{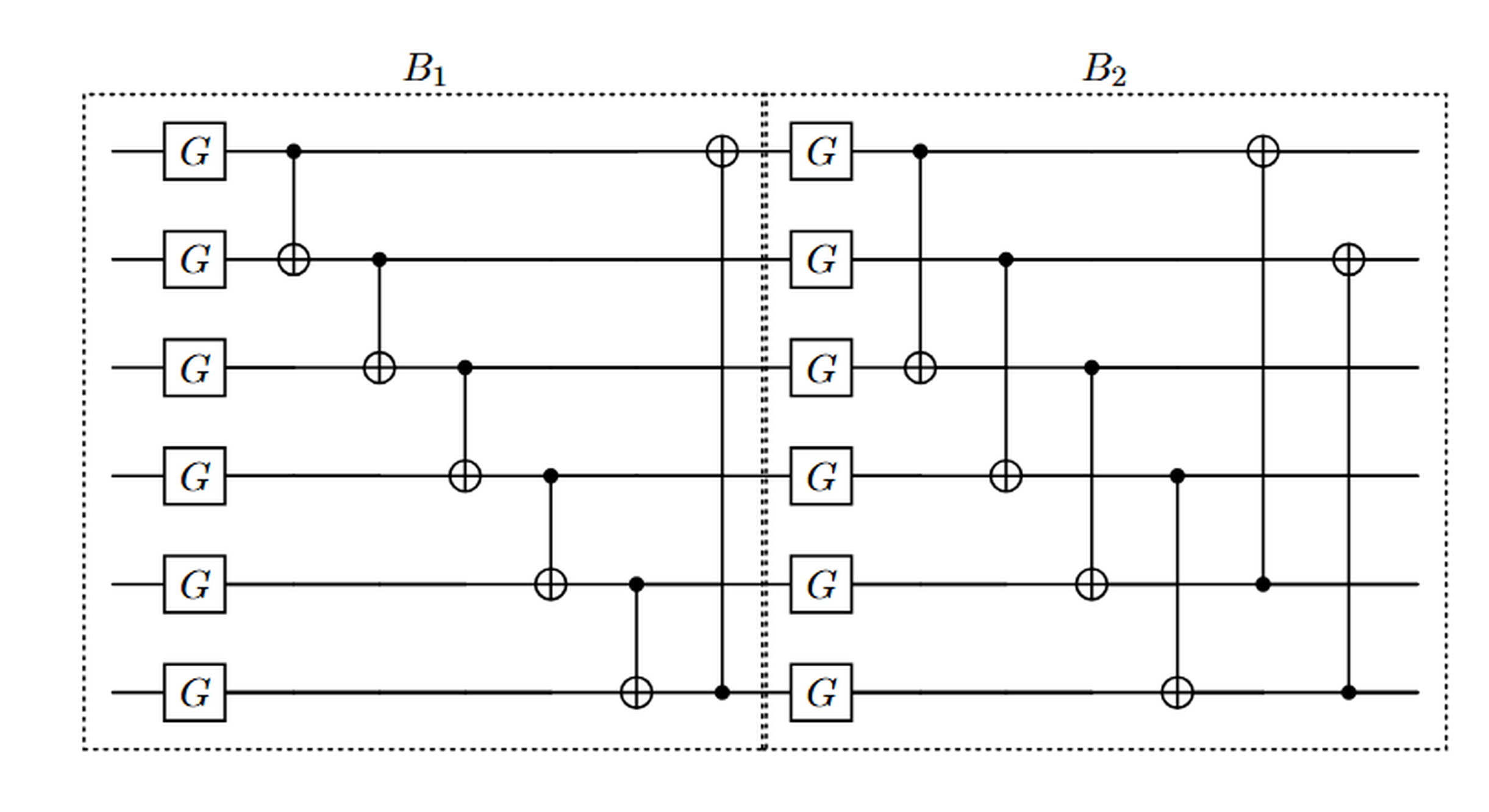}
    \caption{Generic strongly entangling model circuit architecture for 6 qubits. The circuit consists of two \textit{trainable block layers} $B_1$ and $B_2$ with a range of controls of $r=1$ and $r=2$ respectively. The circuit consists of 12 trainable single-qubit gates $G = G(\alpha, \beta, \gamma)$ and 12 controlled single-qubit gates.}
    \label{fig:entangling_circuit}
\end{figure}

We implemented the QNNs mainly with Python version 3.11, Pennylane \cite{bergholm2022pennylaneautomaticdifferentiationhybrid} and JAX \cite{jax2018github}.
Training the models was performed using a full-batch gradient descent approach, where the entire dataset was treated as a single batch for each epoch. For that, we used the pennylanes \textit{default.qubit} simulation device. We used the Mean Squared Error (MSE) as our loss function.
Further, pennylanes \textit{StronglyEntanglingLayers} with 5 entangling layers were used as encoding unitaries $W_{\btheta}^{(l)}$. An example circuit for the strongly entangling circuit can be found in Figure~\ref{fig:entangling_circuit}. 
With that, the number of trainable parameters of the model is given by
\begin{equation}
    n_{\text{trainable parameters}} = R \cdot (L+1) \cdot \text{(\# entangling layers)} \cdot 3 = 15 \cdot (A + R).
\end{equation}
The training process was stopped after 3000 epochs with a learning rate of $0.05$.

\subsection{Results}
\begin{figure}
    \centering
    \includegraphics[width=1.0\linewidth]{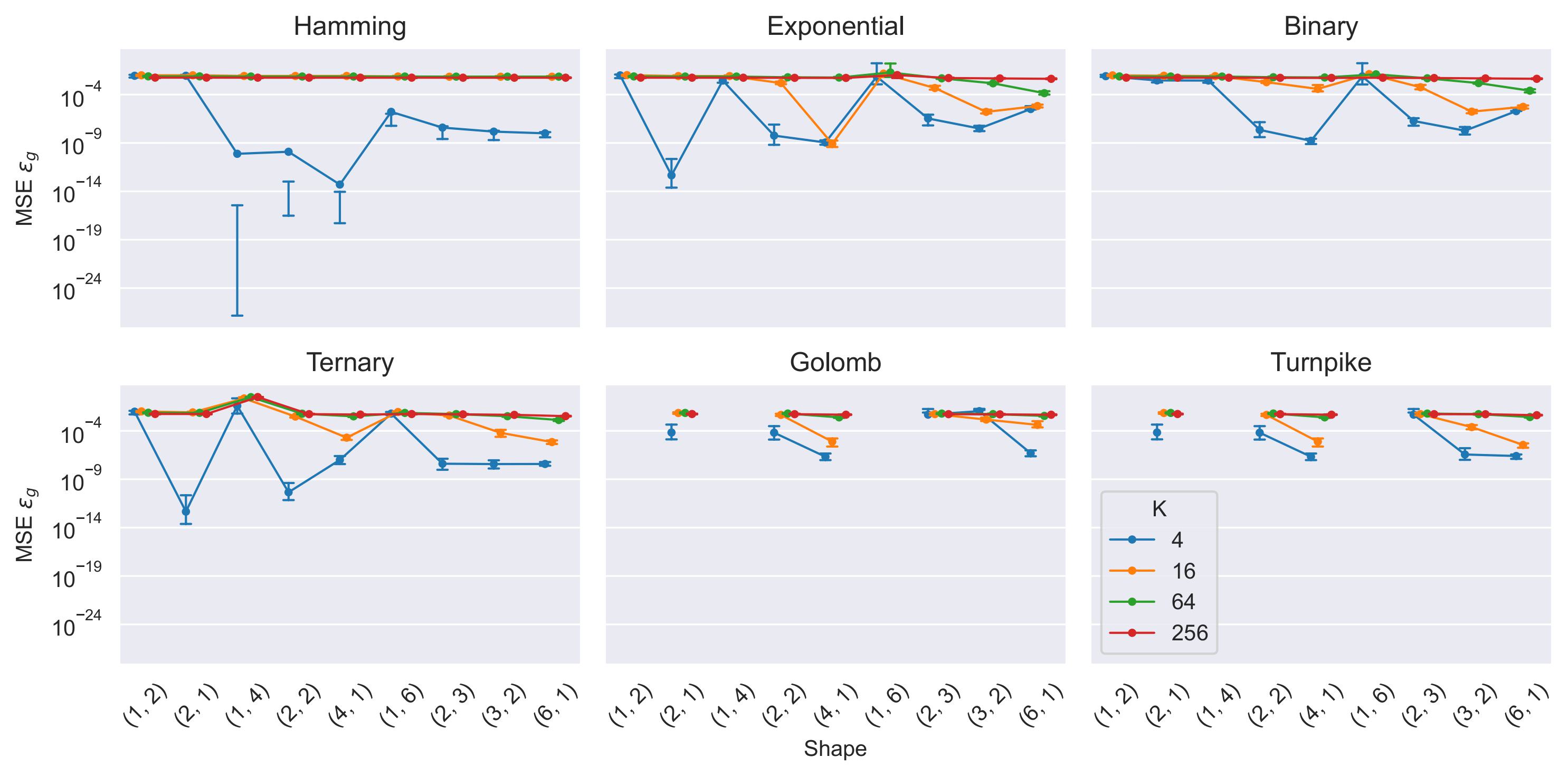}
    \caption{The learning capability defined as the mean of all runs together with the 50\% percentile of all 100 runs with $g \in G_K$. }
    \label{fig:learning capability}
\end{figure}

The results of the experiments can be found in Figure~\ref{fig:learning capability}. 
For $K=4$, only the frequency spectrum of Hamming and binary encoding according to Table~\ref{tab:frequency-spectra} is too small to learn the random Fourier series even theoretically for $A=2$. This is confirmed numerically by the experiment. For $A > 2$, we observe the same effect for all encodings: Obviously, models with $R=1$ are not able to train sufficiently well and to approximate the Fourier series. For increasing $R$ with constant area $A$, the learning ability seems to decrease and thus improve, but at the other extreme it gets slightly worse again at $L=1$, except for Golomb and turnpike. For Hamming with $A=4$, the learning capability is always outside the 50\% percentile, i.e. of the 100 random test functions $g$, there are more than 50 in which the QNN does not learn the function $g$ well, while in the other cases it can obviously learn the function very well. Another effect that can be observed is that the learning capability is better with a smaller area $A$ than with a larger one, provided that the frequency spectrum is large enough to learn the functions well at all. All of the above observations also apply to $K = 16, 64$, provided that the frequency spectrum of the respective encoding is large enough in the respective configuration. However, the learning capability is significantly higher and therefore worse than in the case of $K=4$, which means that the encodings can no longer approximate the test functions nearly as well. For $K=256$, no encoding is able to sufficiently approximate the randomly generated test functions, although the frequency spectrum is theoretically large enough for ternary, Golomb and turnpike.
Compared to each other, the learning capability of encodings with the same shape is better where the frequency spectrum is theoretically smaller, but still covers all frequencies of the randomly generated test function. This is also consistent with the theoretical consideration that the Fourier coefficients have more flexibility there due to the many frequency doublings and are therefore easier to train.

The numerical results of the experiments can be found in Table~\ref{tab:learning-capability} in Appendix~\ref{app:learning-capability}.
In conclusion, we can draw two rules of thumb from these experiments:
\begin{enumerate}
    \item \textbf{Choose the area and encoding so that the frequency spectrum is as large as necessary, but as small as possible.}
    \item \textbf{If the area $A$ is fixed, select the shape $(R, L)$ with the given area so that $R$ is large, but maybe avoid the edge case $(A, 1)$.}
\end{enumerate}

\section{Fischertechnik Fabrik Dataset}
\label{sec:fischertechnik-dataset}
\subsection{Dataset Description}
A real-world dataset was provided by formerly \textit{Umlaut SE}, which is now part of Accenture GmbH specializing in digital engineering and manufacturing solutions. The dataset originates from a simulated industrial process implemented on a \textit{Fischertechnik} factory model. In this setup, the electric current of a pneumatic component was continuously monitored, generating time-series data that reflect the temporal dynamics of the system under normal and fault-induced conditions. The dataset is not publicly available.

The dataset consists of \(479 101\) sensor measurements of electric current, sampled at regular time intervals. During controlled experimental trials, artificial leaks were introduced into the pneumatic system, with precise timestamps of these leak events recorded alongside the sensor readings, providing labeled ground truth for supervised learning tasks. To prepare the data for model training, a sliding window approach was used with a window size of 10 and a stride of 5, extracting temporal features from the time series. Additionally, a Gaussian filter was applied to smooth the current and power measurements, reducing noise and enhancing signal clarity. This was implemented using a one-dimensional Gaussian filter with a standard deviation $\sigma$ of 100, applied to both the current and power time series. This preprocessing yielded a balanced training set with class proportions of \([0.576, 0.424]\), corresponding to normal and leak states. The time-series data were normalized using Min-Max normalization, mapping values to the range \([-1, 1]\) per feature. The data encapsulate real-world signal processing challenges, such as noise (random fluctuations in measurements), transient dynamics (rapid changes during state transitions), and measurement uncertainty (variability due to sensor limitations).

The primary task involves training and validating quantum and classical machine learning models to detect the presence or absence of pneumatic leaks based on the temporal characteristics of the current time series. This requires extracting discriminative features from the raw sensor data and designing algorithms capable of distinguishing subtle differences between normal and leak states.

\begin{figure}[t]
    \centering
    \includegraphics[width=0.9\textwidth]{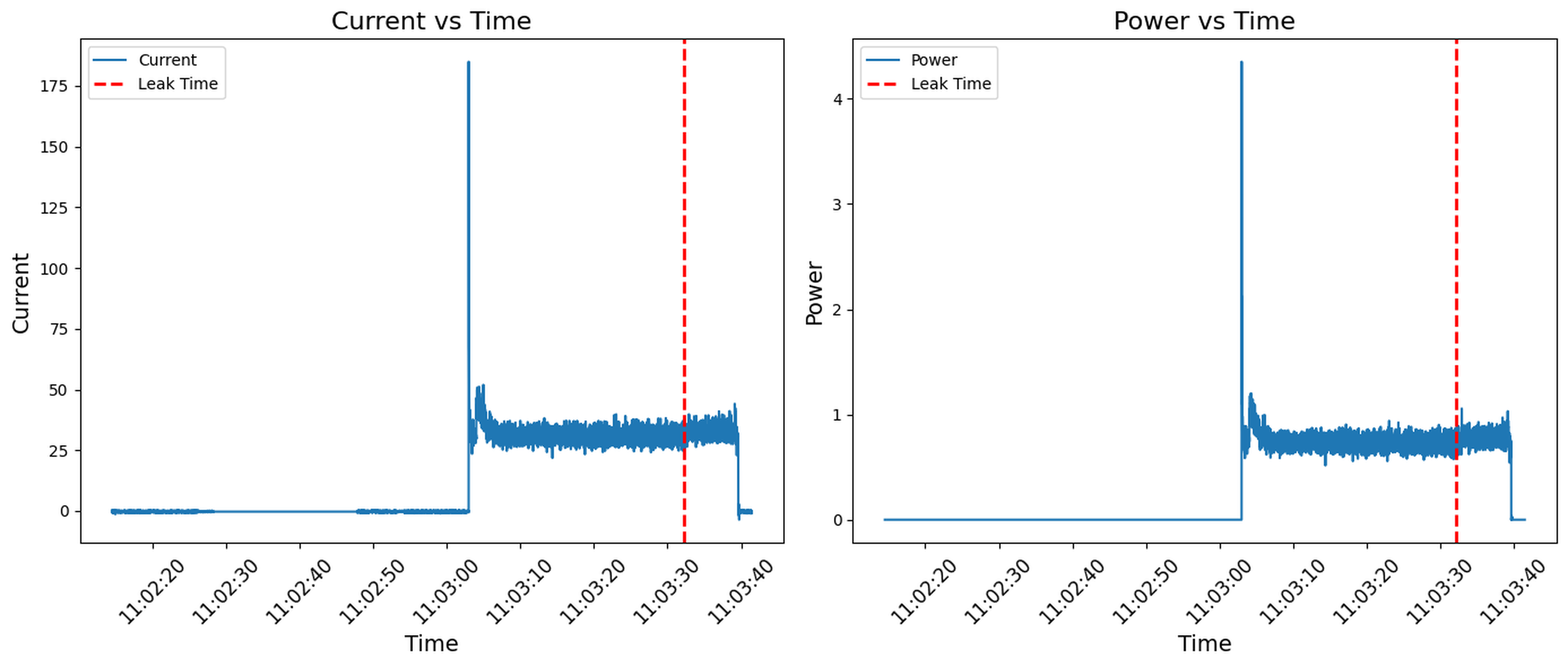}
    \caption{
    Time-series of the Fischertechnik dataset. Left: measured current. Right: corresponding power. The vertical dashed line indicates the onset of a pneumatic leak. Active (nonzero) and inactive (zero) operating periods are visible in both signals.}
    \label{fig:umlaut_dataset}
\end{figure}

Figure~\ref{fig:umlaut_dataset} illustrates key characteristics of the Fischertechnik Fabrik Dataset, displaying time-series visualizations of current and power measurements, with a red dashed line marking the introduction of a pneumatic leak. Operational states—active (nonzero) and inactive (zero)—are discernible in both signals, highlighting the dataset’s utility for fault detection.

Accurate leak detection holds practical significance in industrial automation, enabling early fault identification to reduce energy waste, maintenance costs, and operational downtime.

\subsection{Model}
The performance of Quantum Neural Networks was systematically evaluated to quantify the impact of architectural components.
We reused our implementation of QNNs as described in Section~\ref{subsec:learning-capability-model}. 

For the Fischertechnik dataset, we followed the same area constraints \(A = 2, 4, 6\) as introduced in Section~\ref{subsec:learning-capability-model}. Consequently, for each value of \(A\), all possible shape \((R, L)\) combinations satisfying \(R \times L = A\) were considered. For example, for \(A = 6\), the following shapes were used: \((1,6)\), \((2,3)\), \((3,2)\), and \((6,1)\). The Hamming, binary, exponential, ternary encoding schemes were evaluated under these configurations. 

The quantum neural network (QNN) was trained on a time-series dataset comprising 47 575 samples, each represented as a vector \(\mathbf{x} \in \mathbb{R}^{10}\) (\(N=10\)). These samples were generated through a sliding window procedure (\(window = 10\), \(stride = 5\)) applied to 479 101 raw sensor measurements followed by data cleaning, Gaussian smoothing and normalization. The dataset is split into a training set of 38,060 samples and a test set of 9,515 samples, ensuring a robust evaluation of the model's performance. To mitigate the effects of class imbalance, the training data is balanced to achieve class proportions of approximately \([0.576, 0.424]\)for the normal (y=0) and leak (y=1) classes, respectively. This balancing step is critical for training a QNN that generalizes well across both classes. 

In our approach, we employed a sequential ansatz to decompose the multivariate input into a series of univariate models. Each parameterized unitary $W_{\btheta}^{(l)}$ within the ansatz architecture incorporated 5 strongly entangling layers.

The number of trainable parameters in the sequential ansatz is given by
\begin{equation}
    n_{\text{trainable parameters}} = R \cdot (L \cdot 10 + 1) \cdot 5 \cdot 3 = (A \cdot 10 + R)\cdot 5 \cdot 3.
\end{equation}

Since we have a binary classification task here, the binary cross entropy loss was employed, given by:
\begin{equation}
    \mathrm{BCE} = -\frac{1}{|X_B|} \sum_{i=1}^{|X_B|} \left[y_i \log(\hat{y}_i) + (1 - y_i) \log(1 - \hat{y}_i)\right].
\end{equation}
As our QNN returns values in the range $[-1, 1]$, we mapped the output 
via  $\hat{y}_i = \sigma(6z_i)$ to the range $(0, 1)$, where \( z_i \) is the raw output from the QNN and $\sigma$ the usual sigmoid function. The value $6$ was chosen such that the sigmoid function is saturated enough by the output.

The model was trained using mini-batch gradient descent with a batch size of 64, processing the dataset over 3000 epochs with a learning rate of 0.001. This training configuration enabled effective optimization and leveraged the frequency spectrum properties of the Quantum Neural Network (QNN) as outlined in the theoretical framework.

The model was trained using a virtual machine on Google Cloud Platform (GCP) configured with the C2D-standard-8 machine type. This setup featured 8 virtual CPUs (4 physical cores) and 32 GB of memory, powered by AMD EPYC Milan processors. Depending on the specific QNN architecture (values of $R$, $L$, and encoding scheme), the training time for 3000 epochs for an area constraint of $A=4$ ranged from $\approx$39 minutes (e.g., for $R=1, L=4$ with exponential encoding) to $\approx$5.1 hours (e.g., for $R=4, L=1$ with hamming encoding). For context, with a larger area constraint of $A=6$, these times extended from $\approx$48 minutes to $\approx$28.5 hours.

The performance of the model was evaluated using the accuracy, precision, F1 score, and ROC AUC metrics.

\subsection{Results}
\begin{sidewaystable}
\caption{Performance metrics on the Fischertechnik Fabrik dataset for QNNs with different encodings, areas $A$, and shapes $(R,L)$.}
\label{tab:qnn-fischertechnik-metrics}
\centering
\begin{tabular*}{\textheight}{@{\extracolsep\fill}lrcrrrrrrr}
\toprule
\multirow{2}{*}{Encoding} & \multirow{2}{*}{Area} & \multirow{2}{*}{Shape} &
\multirow{2}{*}{$n_{\text{train}}$} & \multicolumn{6}{c}{Performance metrics} \\
\cmidrule(lr){5-10}
 &  &  &  & Loss & Accuracy & Precision & Recall & F1-Score & ROC-AUC \\
\midrule
Hamming     & 2 & (1, 2) & 315 & 0.3402 & 0.8687 & 0.7853 & 0.9502 & 0.8599 & 0.9006 \\
            &   & (2, 1) & 330 & 0.3337 & 0.8644 & 0.7864 & 0.9338 & 0.8538 & 0.9202 \\
            & 4 & (1, 4) & 615 & 0.3463 & 0.8707 & 0.7838 & 0.9598 & 0.8629 & 0.9200 \\
            &   & (2, 2) & 630 & 0.3189 & 0.8679 & 0.7766 & 0.9663 & 0.8612 & 0.9202 \\
            &   & (4, 1) & 660 & 0.3357 & 0.8698 & 0.7822 & 0.9603 & 0.8621 & 0.8797 \\
            & 6 & (1, 6) & 915 & 0.2959 & 0.8660 & 0.7782 & 0.9598 & 0.8595 & 0.9297 \\
            &   & (2, 3) & 930 & 0.2950 & 0.8657 & 0.7779 & 0.9592 & 0.8592 & 0.9294 \\
            &   & (3, 2) & 945 & 0.2951 & 0.8657 & 0.7775 & 0.9601 & 0.8597 & 0.9302 \\
            &   & (6, 1) & 990 & 0.2986 & 0.8658 & 0.7811 & 0.9528 & 0.8585 & 0.9292 \\
\midrule
Exponential & 2 & (1, 2) & 315 & 0.3109 & 0.8660 & 0.7770 & 0.9593 & 0.8586 & 0.9292 \\
            &   & (2, 1) & 330 & 0.3103 & 0.8639 & 0.7914 & 0.9219 & 0.8517 & 0.9273 \\
            & 4 & (1, 4) & 615 & 0.3218 & 0.8701 & 0.7899 & 0.9550 & 0.8665 & 0.9267 \\
            &   & (2, 2) & 630 & 0.3182 & 0.8708 & 0.7894 & 0.9578 & 0.8670 & 0.9299 \\
            &   & (4, 1) & 660 & 0.3182 & 0.8728 & 0.7925 & 0.9567 & 0.8681 & 0.9306 \\
            & 6 & (1, 6) & 915 & 0.2935 & 0.8641 & 0.7765 & 0.9524 & 0.8560 & 0.9287 \\
            &   & (2, 3) & 930 & 0.2863 & 0.8670 & 0.7889 & 0.9408 & 0.8578 & 0.9318 \\
            &   & (3, 2) & 945 & 0.2872 & 0.8712 & 0.7927 & 0.9411 & 0.8600 & 0.9345 \\
            &   & (6, 1) & 990 & 0.2807 & 0.8671 & 0.7905 & 0.9399 & 0.8579 & 0.9317 \\
\midrule
Binary      & 2 & (1, 2) & 315 & 0.3299 & 0.8445 & 0.7914 & 0.8597 & 0.8241 & 0.9245 \\
            &   & (2, 1) & 330 & 0.3181 & 0.8647 & 0.7858 & 0.9349 & 0.8541 & 0.9246 \\
            & 4 & (1, 4) & 615 & 0.3184 & 0.8714 & 0.7909 & 0.9455 & 0.8618 & 0.9279 \\
            &   & (2, 2) & 630 & 0.3003 & 0.8777 & 0.7990 & 0.9488 & 0.8676 & 0.9384 \\
            &   & (4, 1) & 660 & 0.3123 & 0.8771 & 0.7929 & 0.9570 & 0.8683 & 0.9316 \\
            & 6 & (1, 6) & 915 & 0.2928 & 0.8652 & 0.7811 & 0.9528 & 0.8585 & 0.9292 \\
            &   & (2, 3) & 930 & 0.2875 & 0.8671 & 0.7904 & 0.9387 & 0.8582 & 0.9322 \\
            &   & (3, 2) & 945 & 0.2802 & 0.8711 & 0.7931 & 0.9417 & 0.8603 & 0.9351 \\
            &   & (6, 1) & 990 & 0.2859 & 0.8671 & 0.7902 & 0.9401 & 0.8580 & 0.9319 \\
\midrule
Ternary     & 2 & (1, 2) & 315 & 0.3109 & 0.8660 & 0.7770 & 0.9593 & 0.8586 & 0.9292 \\
            &   & (2, 1) & 330 & 0.3103 & 0.8639 & 0.7914 & 0.9219 & 0.8517 & 0.9273 \\
            & 4 & (1, 4) & 615 & 0.6943 & 0.7575 & 0.7317 & 0.6760 & 0.7027 & 0.8323 \\
            &   & (2, 2) & 630 & 0.4502 & 0.8170 & 0.7679 & 0.8146 & 0.7906 & 0.9056 \\
            &   & (4, 1) & 660 & 0.4502 & 0.8170 & 0.7679 & 0.8146 & 0.7906 & 0.9056 \\
            & 6 & (1, 6) & 915 & 0.4749 & 0.8143 & 0.7200 & 0.8515 & 0.7800 & 0.8900 \\
            &   & (2, 3) & 930 & 0.3034 & 0.8610 & 0.7762 & 0.9528 & 0.8555 & 0.9190 \\
            &   & (3, 2) & 945 & 0.2737 & 0.8666 & 0.7881 & 0.9402 & 0.8575 & 0.9328 \\
            &   & (6, 1) & 990 & 0.3005 & 0.8665 & 0.7879 & 0.9405 & 0.8575 & 0.9328 \\
\bottomrule
\end{tabular*}
\footnotesize
\textit{Note.} $A$ denotes the area of the QNN, defined as $A = R \times L$ with $R$ the number of qubits and $L$ the number of layers. The shape $(R,L)$ specifies the corresponding architecture, and $n_{\text{train}}$ is the number of trainable parameters. All models were trained for 3000 epochs with a learning rate of $0.001$ on the Fischertechnik dataset.
\end{sidewaystable}

Table~\ref{tab:qnn-fischertechnik-metrics} summarizes the QNN performance on the Fischertechnik dataset across encoding families and architectural configurations
$(R,L)$ under a fixed area $A = R \times L$ and learning rate $\eta = 0.001$. 

In the $A=6$ regime, binary and exponential provide the most favorable operating point, reaching accuracies of
$0.8711$ and $0.8712$ at $(R,L)=(3,2)$ with low losses ($0.2802$-$0.2872$).
In contrast, the single-qubit shape $(1,6)$ slightly degrades for these encodings (accuracy $0.8641$-$0.8652$),
which is qualitatively consistent with the learning capability results indicating that $R=1$ configurations become harder to train once $A>2$, despite spectral invariance at fixed $A$.

Importantly, increasing area is not monotonically beneficial: for binary, $A=4$ already attains $\approx 0.877$
accuracy (e.g., $(2,2)$ and $(4,1)$), suggesting that the effective task bandwidth is covered at smaller $A$ and
that further spectral expansion does not translate into improved class separability.
Ternary provides a particularly clear instance of the expressivity-trainability trade-off: although it offers a theoretically maximal spectrum, the $A=4$ configurations and the deep single-qubit case $(1,6)$ at $A=6$ underperform markedly (accuracy $0.7575$-$0.8170$ and $0.8143$, loss $0.4502$-$0.6943$ and $0.4749$),
whereas distributing the same area across more qubits at $A=6$ ($(2,3)$, $(3,2)$, $(6,1)$) recovers competitive
performance (accuracy $\approx 0.8610$--$0.8666$, loss $0.2737$-$0.3034$). Finally, Hamming, consistent with its restricted spectrum, remains comparatively stable but exhibits a higher final loss level ($\approx 0.295$ at $A=6$) without accuracy gains over moderately richer encodings.

\begin{figure}[t]
  \centering
  \includegraphics[width=\linewidth]{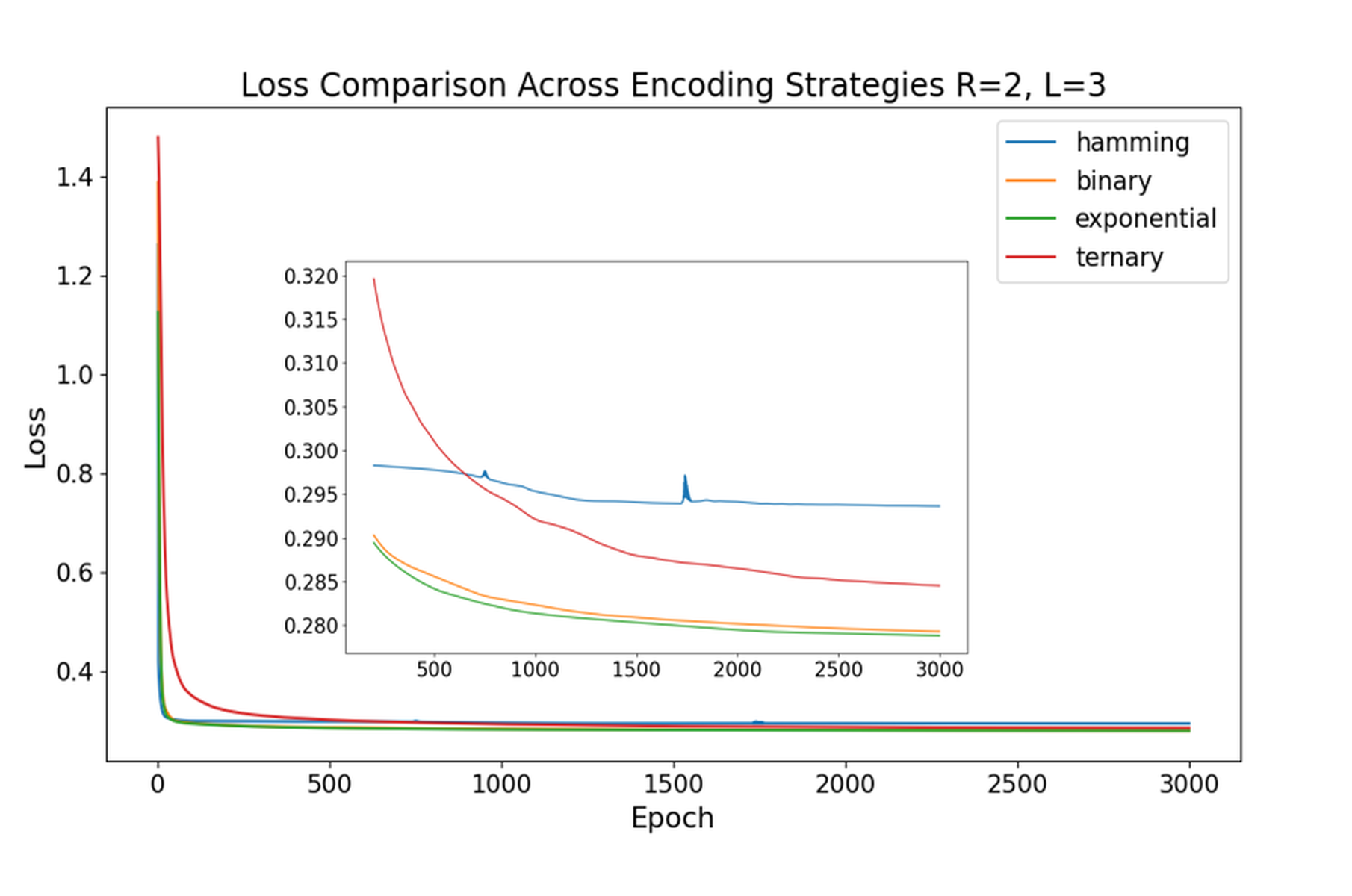}
  \caption{Training loss versus epoch for four encoding strategies (Hamming, binary, exponential, and ternary) with $R=2$ and $L=3$ ($A=6$), trained for 3000 epochs at learning rate $\eta=0.005$ (other hyperparameters fixed). Inset: zoom of the late-epoch region to highlight differences in the converged loss values and transient behavior.}
  \label{fig:loss_comparison_encodings_R2L3}
\end{figure}

Figure~4 shows the training loss trajectories on the Fischertechnik dataset for the fixed-area configuration $(R,L)=(2,3)$ ($A=6$), trained for 3000 epochs at learning rate $\eta = 0.005$ (all other hyperparameters fixed).
All encodings show a steep initial loss decrease followed by a slower convergence regime. The inset highlights consistent late-epoch differences in the converged loss levels and in transient behavior.
In particular, binary and exponential converge smoothly to the lowest loss plateaus, whereas Hamming stabilizes at a higher loss level and exhibits transient irregularities late in training. Ternary shows a slower relaxation toward its asymptotic regime, consistent with a more challenging optimization landscape despite its potentially broader spectrum.

Although Fig.~4 focuses on $(R,L)=(2,3)$, auxiliary runs for the single-qubit shape $(R,L)=(1,6)$ (not shown) indicate that Hamming can exhibit pronounced loss oscillations at $\eta = 0.005$ across random initializations.
This behavior is consistent with Hamming's constrained spectrum $\Omega = \mathbb{Z}_6=\{-6,\ldots,6\}$ and the associated degeneracy effects, which can induce a more irregular loss landscape \cite{holzer2024specinv}.
In these auxiliary experiments, reducing the learning rate to $\eta = 0.001$ largely suppresses the oscillations, whereas allocating the same area across more qubits (e.g., $(R,L)=(2,3)$) yields smoother and more stable convergence even at $\eta = 0.005$.

Taken together, the empirical ordering observed at $A=6$ with binary and exponential providing the most reliable operating point in terms of low loss and stable convergence, followed by ternary, and with Hamming being most sensitive to step size, is consistent with the learning capability heuristics from Section~\ref{sec:learning-capability}:
\begin{enumerate}
    \item richer, less-degenerate spectra improve expressiveness and sample efficiency up to the point where the task bandwidth is covered;
      \item distributing capacity across qubits ($R>1$ at fixed area) smooths optimization and improves stability;
  \item restricted spectra can amplify step-size sensitivity, explaining oscillatory behavior at larger $\eta$ and its mitigation at smaller $\eta$.
\end{enumerate}

In practice, these results suggest prioritizing binary, exponential encodings and, under a fixed area constraint, favoring multi-qubit shapes to obtain stable, low-loss training.

The extended results are provided in the Appendix~\ref{app:fischertechnik} for further reference.

\section{NASA Bearing Dataset}
\label{sec:nasa-bearing-dataset}
\subsection{Dataset Description}

\begin{figure}[ht]
    \centering \includegraphics[width=0.5\textwidth]{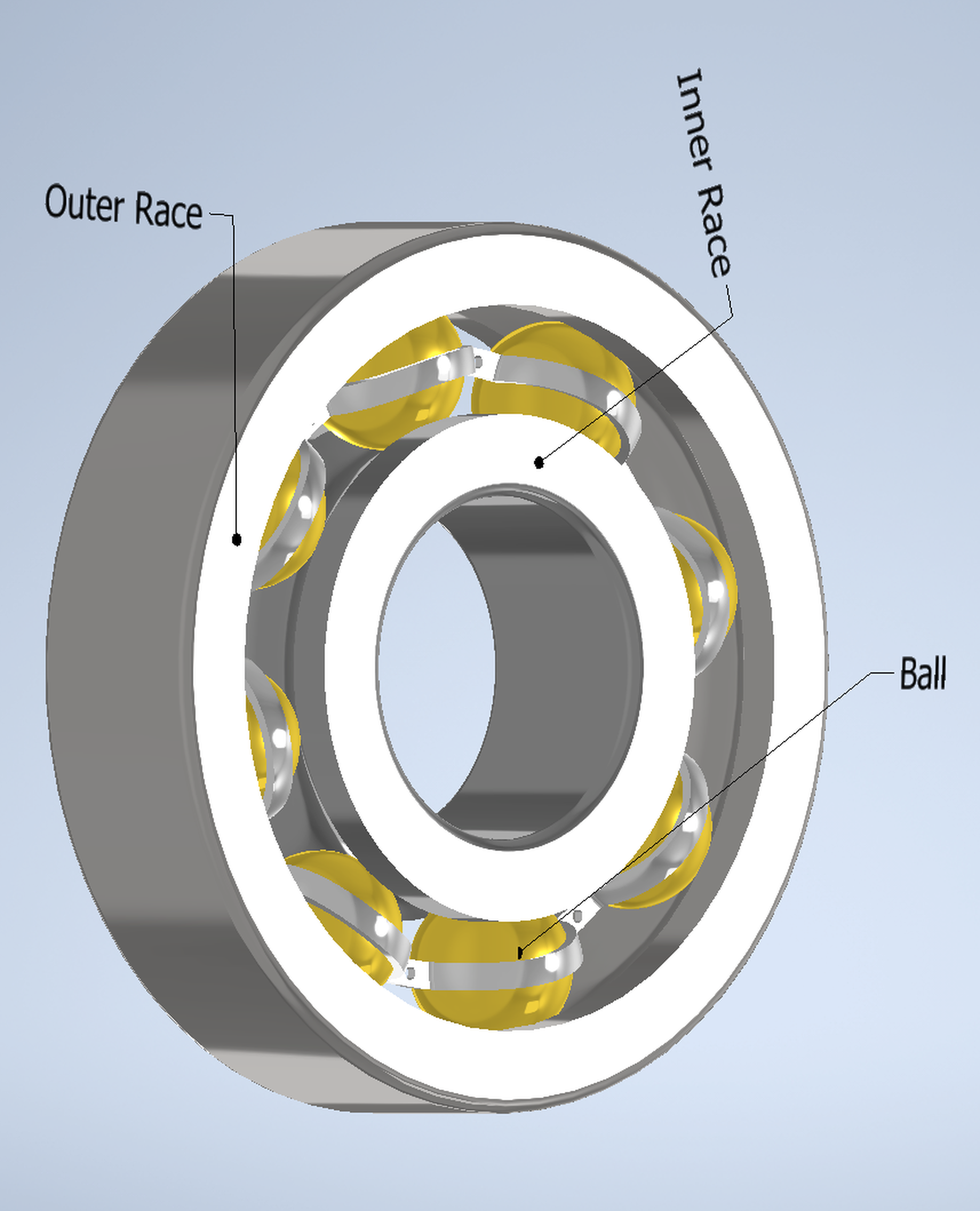}
    \caption{A simplified depiction of a ball bearing. For the second NASA dataset, the outer race of bearing 1 failed.}
    \label{fig:ball_bearing}
\end{figure}

The third dataset, herein referred to as the \textit{NASA} dataset, was sourced from the Kaggle repository's NASA Bearing Dataset \cite{lee2007bearing} section. Vibration data were collected using LabVIEW in conjunction with the DAQCard-6062E data acquisition card. The experiment utilized four Rexnord ZA-2115 double row bearings, each comprising 16 rollers per row, with a pitch diameter of 2.815 inches, roller diameter of 0.331 inches, and a tapered contact angle of 15.17\textdegree. Accelerometers were installed on each bearing housing to measure the bearing vibration signals, while four thermocouples were attached to the outer race of each bearing to monitor bearing temperature. A simplified schematic of the bearing components, including the outer race referenced throughout this section, is shown in Figure~\ref{fig:ball_bearing}.

The bearings were mounted on a shaft that rotated at a constant 2000 RPM driven by an AC motor connected to the shaft via a rub belt. A 6000 lbs radial load was applied to the shaft and bearing with the use of a spring mechanism. In order to ensure controlled lubrication, all four bearings were forced lubricated by an oil circulation system that regulated the flow and temperature of the lubricant. Magnetic plugs were installed in the system to collect any debris as evidence of bearing degradation and facilitate efficient maintenance. The experiment concluded once the amount of debris collected by the plugs exceeded a certain threshold, indicating significant bearing wear. 

\begin{figure}[ht]
\centering
\includegraphics[width=1.0\textwidth]{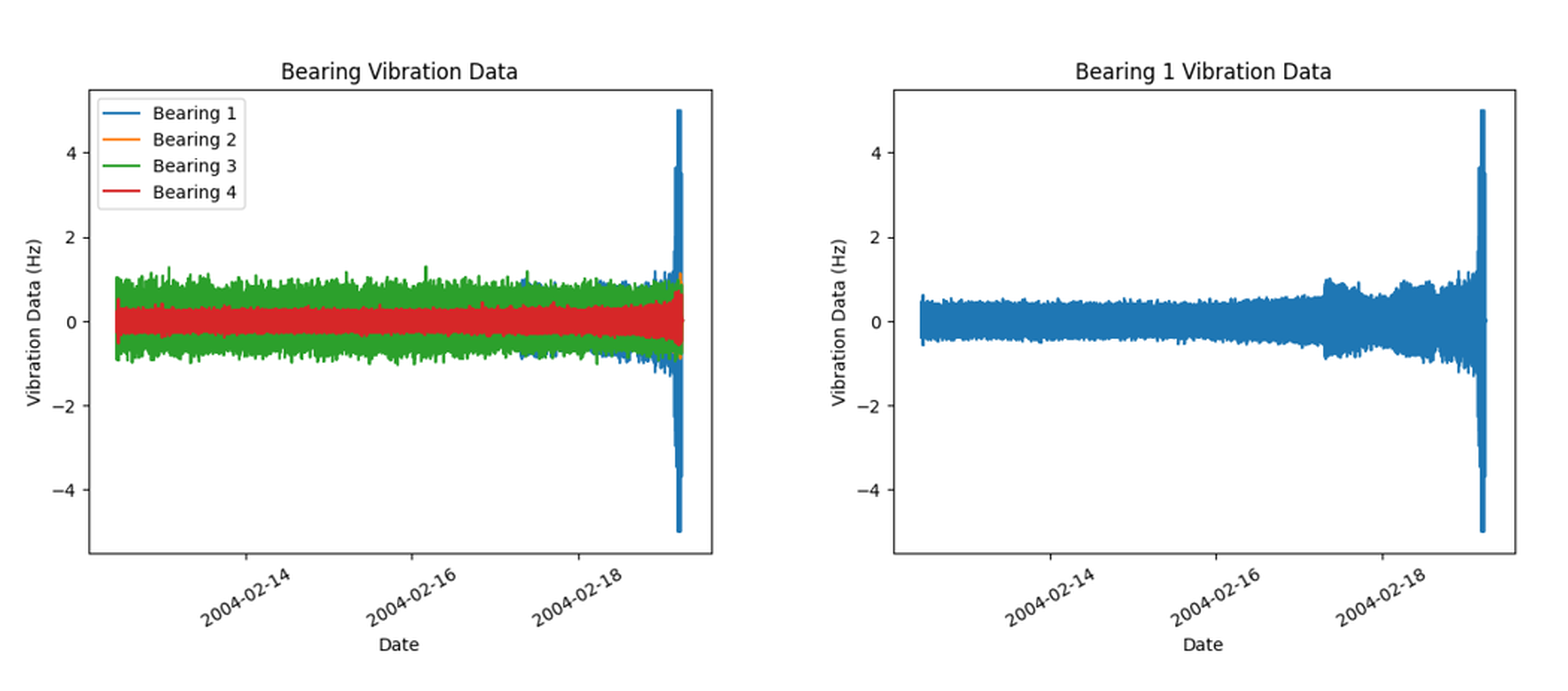}
\caption{Progression of bearing health in the second NASA dataset, visualized via RMS vibration values. Left: Comparative RMS trends for all four bearings. Right: Detailed RMS plot for Bearing 1, showing increasing vibration levels indicative of an outer race failure. This dataset was employed for QNN model training.}
\label{fig:bearing-vibration-combined}
\end{figure}

The dataset consists of four independent test-to-failure runs. In the second run, which was used to train the Quantum Neural Network (QNN), bearing~1 developed an outer-race fault. Vibration signals were recorded from four bearings from February~12,~2004 10{:}32{:}39 to February~19,~2004 06{:}22{:}39. Every 10 minutes, the system captured a one-second snapshot from each bearing channel (20{,}480 samples at 20.48~kHz), yielding 984 snapshots in total. Accordingly, the raw measurements can be represented as a three-dimensional array of shape $(984, 4, 20480)$, where the axes correspond to snapshot index, bearing channel, and time sample within each snapshot. The evolution of the vibration signal over the run is shown in Figure~\ref{fig:bearing-vibration-combined}. 

To prepare the data for QNN training, we applied a two-stage preprocessing pipeline. In the first stage, features were extracted from the raw vibration data. For each of the 984 one-second snapshots, the root mean square (RMS) values were computed for the vibration signals of each of the four bearings. The RMS value for a bearing from a given 1-second snapshot is defined as:

\begin{equation}
    RMS = \sqrt{\frac{1}{N} \sum_{i=1}^{N}x_i^2}
\end{equation}
where $x_i$ is the vibration signal amplitude at sample $i$, and $N$ is the number of samples in the 1-second time window (20,480 samples). This step produced a matrix of RMS values with the shape (984, 4).
, with each row corresponding to the time point and each column representing the RMS value for one of the four bearings. 

To quantify system-level deviations and aid in subsequent labeling, the Mahalanobis distance was calculated using the RMS values of the four bearings. The Mahalanobis distance, $D_m$, is defined as:
\begin{equation}
    D_m(x) = \sqrt{(x-\mu)^T S^{-1} (x-\mu)}
\end{equation}
where $x$ is the vector of RMS values for the four bearings at a given time, $\mu$ is the mean vector of RMS values derived from a reference set of healthy bearing data, and $S^{-1}$ is the inverse of the covariance matrix of the healthy data. This metric accounts for correlations among the bearings’ RMS values, providing a normalized measure of deviation from the healthy state. The Mahalanobis distance was computed for each of the 984 time points, resulting in a one-dimensional array of shape (984, 1). Each time point was subsequently labeled as healthy (0) or anomalous (1) by applying a threshold to its MD value. This concluded the first stage, yielding a dataset containing the four RMS values, the MD, and a binary label for each of the 984 time points.

In the second stage, this processed dataset was further prepared for input to the QNN. From the features generated in the first stage, only the four RMS values (one for each bearing) were selected as input features for the QNN model. Each of the 984 time points, now represented by a 4-dimensional RMS feature vector and its corresponding label, was treated as an individual sample.
This dataset of approximately (984, 4) features and corresponding labels was then split into training (80\%) and testing (20\%) sets using stratified sampling to maintain similar class proportions in both sets. To address class imbalance in the training data, the Synthetic Minority Over-sampling Technique (SMOTE) was applied exclusively to the training set. This technique generates synthetic samples for the minority class (anomalous instances), creating a more balanced dataset for training the QNN. 

Finally, both the SMOTE-augmented training features and the test features were scaled to a range of [-1, 1] using Min-Max scaling. The scaler was fitted only on the training data and then applied to both training and test sets to prevent data leakage. This sequence resulted in the final datasets used to train and evaluate the QNN, where each input sample to the QNN consisted of four scaled RMS features.

\subsection{Model}

The same QNN model was reused as in Section~\ref{subsec:learning-capability-model}: the sequential ansatz with 5 strongly entangling layers per unitary $W_{\boldsymbol\theta}^{(l)}$, the same encoding families (Hamming, binary, exponential, ternary, Golomb, ternary), the same binary cross-entropy objective with the output mapping $\hat{y}=\sigma(6z)$, and the same training regime (mini-batch size 64, 3000 epochs, learning rate 0.001). The only architectural change is the \emph{input size}, which is $N=4$ here (four per-bearing RMS features) instead of $N=10$ in the Fischertechnik dataset. Preprocessing also differs slightly: for NASA we applied SMOTE on the training split and scaled features to $[-1,1]$, whereas the Fischertechnik pipeline used Gaussian smoothing/normalization without SMOTE. All runs were executed on the same GCP C2D-standard-8 VM (8 vCPUs, 32 GB RAM). Because the dataset is smaller (984 samples), wall-clock training per configuration was substantially shorter, typically on the order of $\sim$2--10 minutes for $A\in{2,4}$ and $\sim$8--20 minutes for $A=6$.

\subsection{Results}
The NASA Dataset provides high resolution vibration data from bearings operated to failure, which we use to assess QNNs for fault detection and remaining useful life (RUL) prediction.

\begin{figure}[t]
    \centering
    \includegraphics[width=1.0\textwidth]{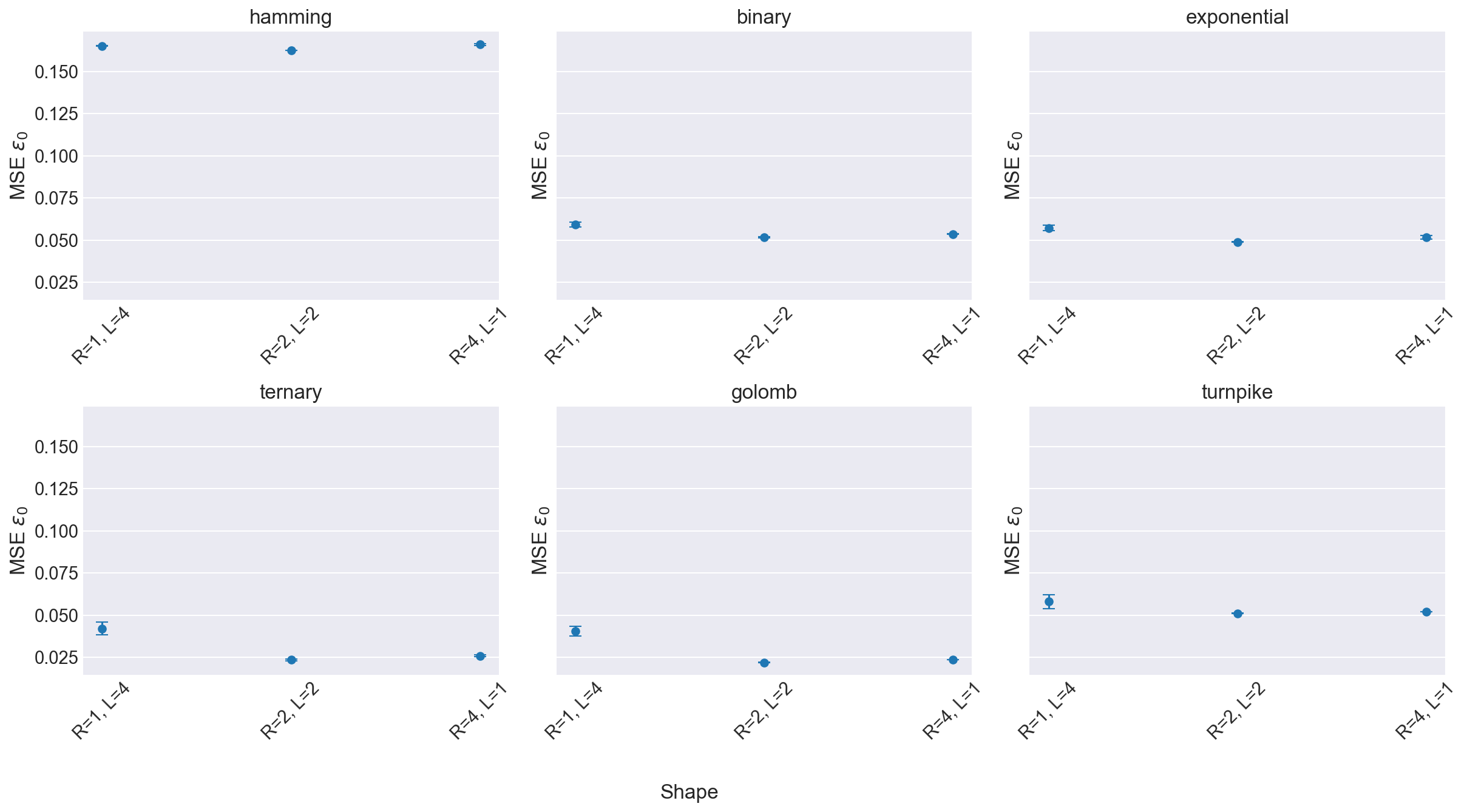}
    \caption{Loss comparison across encoding strategies (Hamming, binary, exponential, ternary, Golomb, turnpike) for different \((R,L)\) configurations with \(A=4\). Each point is the mean loss over three runs and error bars indicate standard deviation.}
    \label{fig:loss_comparison_NASA}
\end{figure}

Figure~\ref{fig:loss_comparison_NASA} shows convergence differences across encoding strategies. With area fixed at \(A=4\), that is, comparing shapes \((R,L)\in\{(1,4),(2,2),(4,1)\}\), the overall convergence level is similar across shapes and the observed differences are primarily attributable to the encoding. Golomb attains the lowest mean loss (0.022), followed by Ternary (0.023 at \(R=2\), \(L=2\)). Hamming exhibits higher baseline losses (\(\approx 0.165\) to \(0.166\)) and greater variance, particularly for \((R,L)=(1,4)\). These observations are consistent with spectral invariance: for fixed area, shape changes have limited effect, whereas the encoding restructures the optimization landscape.

Table~\ref{tab:qnn_performance_NASA} summarizes classification metrics: accuracy up to \(98.90\%\) and ROC AUC near \(1.00\), with consistently strong results for binary, exponential, ternary, Golomb and turnpike. Additional learning rates (0.01 and 0.005) yield similar trends with minor differences in convergence speed (Appendix~\ref{app:nasa}).

Compared to the learning capability study in Section~\ref{sec:learning-capability}, the pattern is consistent. Restricted spectra (Hamming) underperform. Moderately rich spectra with low degeneracy (binary and exponential) train reliably and already saturate ROC-AUC. The largest spectra (ternary, Golomb and turnpike) reduce cross-entropy slightly without notable accuracy gains. Once the effective bandwidth of the task is covered, additional frequencies mainly allow finer adjustment of the trainable Fourier coefficients \(c_{\omega}(\boldsymbol{\theta})\) and of the measured expectation value \(\langle M\rangle\) rather than increase class separability. For fixed \(A=4\), differences between \((1,4)\), \((2,2)\) and \((4,1)\) are small and slightly favor \((2,2)\), which reflects the mild advantage of larger \(R\) at constant area.

Overall, the NASA results support the heuristics from Section~\ref{sec:learning-capability}: select encoding and area so that the spectrum is as large as necessary but not larger, and for fixed area prefer allocating capacity to qubits rather than to depth. In this dataset, binary and exponential are strong defaults. ternary, Golomb and turnpike yield small BCE improvements by enabling finer control over \(c_{\omega}(\boldsymbol{\theta})\) and the measurement signal, whereas Hamming is too restrictive.

\begin{sidewaystable}
\caption{Performance metrics on the NASA bearing dataset for QNNs with different encoding strategies and shapes $(R,L)$.}
\label{tab:qnn_performance_NASA}%
\centering
\begin{tabular*}{\textheight}{@{\extracolsep\fill}lcrrrrrr}
\toprule
\multirow{2}{*}{Encoding} & \multirow{2}{*}{Shape} &
  \multicolumn{6}{c}{Performance metrics} \\
\cmidrule(lr){3-8}
& & Loss & Accuracy & Precision & Recall & F1 Score & ROC-AUC \\
\midrule
Hamming     & (1,4) & $0.165 \pm 0.0001$ & $0.9645 \pm 0.1128$ & $1.0000 \pm 0.0000$ & $0.9079 \pm 0.0000$ & $0.9517 \pm 0.0000$ & $0.9994 \pm 0.0001$ \\
            & (2,2) & $0.163 \pm 0.0000$ & $0.9670 \pm 0.0000$ & $1.0000 \pm 0.0000$ & $0.9145 \pm 0.0000$ & $0.9553 \pm 0.0000$ & $0.9993 \pm 0.0000$ \\
            & (4,1) & $0.166 \pm 0.0006$ & $0.9645 \pm 0.1302$ & $1.0000 \pm 0.0000$ & $0.9079 \pm 0.0000$ & $0.9517 \pm 0.0000$ & $0.9995 \pm 0.0000$ \\
\midrule
Binary      & (1,4) & $0.0593 \pm 0.0014$ & $0.9755 \pm 0.0000$ & $0.9931 \pm 0.1196$ & $0.9430 \pm 0.0031$ & $0.9674 \pm 0.0016$ & $0.9921 \pm 0.0028$ \\
            & (2,2) & $0.0520 \pm 0.0002$ & $0.9831 \pm 0.0651$ & $0.9932 \pm 0.4314$ & $0.9627 \pm 0.0112$ & $0.9777 \pm 0.0058$ & $0.9987 \pm 0.0008$ \\
            & (4,1) & $0.0540 \pm 0.0002$ & $0.9865 \pm 0.0651$ & $0.9933 \pm 0.2393$ & $0.9715 \pm 0.0062$ & $0.9823 \pm 0.0032$ & $0.9997 \pm 0.0000$ \\
\midrule
Exponential & (1,4) & $0.0572 \pm 0.0015$ & $0.9797 \pm 0.1723$ & $0.9932 \pm 0.2072$ & $0.9539 \pm 0.0054$ & $0.9766 \pm 0.0028$ & $0.9926 \pm 0.0054$ \\
            & (2,2) & $0.0490 \pm 0.0003$ & $0.9848 \pm 0.0000$ & $0.9932 \pm 0.2072$ & $0.9671 \pm 0.0054$ & $0.9800 \pm 0.0028$ & $0.9968 \pm 0.0034$ \\
            & (4,1) & $0.0520 \pm 0.0010$ & $0.9865 \pm 0.0651$ & $0.9933 \pm 0.1196$ & $0.9715 \pm 0.0031$ & $0.9823 \pm 0.0016$ & $0.9997 \pm 0.0000$ \\
\midrule
Ternary     & (1,4) & $0.0421 \pm 0.0038$ & $0.9628 \pm 0.2604$ & $0.9835 \pm 0.011784$ & $0.9189 \pm 0.0265$ & $0.9499 \pm 0.0165$ & $0.9907 \pm 0.0050$ \\
            & (2,2) & $0.0230 \pm 0.0005$ & $0.9848 \pm 0.0000$ & $0.9978 \pm 0.7179$ & $0.9627 \pm 0.0203$ & $0.9798 \pm 0.0098$ & $0.9976 \pm 0.0017$ \\
            & (4,1) & $0.0260 \pm 0.0007$ & $0.9890 \pm 0.0000$ & $0.9955 \pm 0.5215$ & $0.9759 \pm 0.0112$ & $0.9856 \pm 0.0069$ & $0.9994 \pm 0.0003$ \\
\midrule
Golomb      & (1,4) & $0.0400 \pm 0.0030$ & $0.9712 \pm 0.1723$ & $0.9839 \pm 1.0634$ & $0.9408 \pm 0.0234$ & $0.9617 \pm 0.0146$ & $0.9903 \pm 0.0062$ \\ 
            & (2,2) & $0.0220 \pm 0.0002$ & $0.9848 \pm 0.0000$ & $0.9955 \pm 0.4145$ & $0.9649 \pm 0.0112$ & $0.9799 \pm 0.0056$ & $0.9989 \pm 0.0008$ \\
            & (4,1) & $0.0240 \pm 0.0001$ & $0.9882 \pm 0.0000$ & $0.9977 \pm 0.2393$ & $0.9715 \pm 0.0031$ & $0.9844 \pm 0.0031$ & $0.9994 \pm 0.0002$ \\
\midrule
Turnpike    & (1,4) & $0.0581 \pm 0.0042$ & $0.9822 \pm 0.0651$ & $0.9932 \pm 0.2072$ & $0.9605 \pm 0.0054$ & $0.9766 \pm 0.0028$ & $0.9952 \pm 0.0054$ \\
            & (2,2) & $0.0511 \pm 0.0001$ & $0.9839 \pm 0.0000$ & $0.9932 \pm 0.3166$ & $0.9649 \pm 0.0082$ & $0.9788 \pm 0.0043$ & $0.9991 \pm 0.0008$ \\
            & (4,1) & $0.0520 \pm 0.0001$ & $0.9865 \pm 0.0000$ & $0.9933 \pm 0.1196$ & $0.9715 \pm 0.0031$ & $0.9823 \pm 0.0016$ & $0.9996 \pm 0.0002$ \\
\bottomrule
\end{tabular*}

\footnotesize
\textit{Note.} All models were trained for 3000 iterations with a learning rate of $0.001$. Reported loss, accuracy, precision, recall, F1 score, and ROC-AUC are means $\pm$ standard deviations over three runs on the NASA bearing dataset.
\end{sidewaystable}

\section{Conclusion and Outlook} 
\label{sec:conclusion}

This work systematically evaluated Quantum Neural Networks across synthetic and real-world datasets, studying the effect of encoding strategies and architectural configurations like the shape, area and sequential or parallel ansatz. 

We measured the learning capability of the models with all possible configurations on randomly generated finite Fourier series with different frequency ranges. First, while in theory the frequency spectrum of the encodings depends only on the area $A$ and not on the shape $(R, L)$, the results show a strong dependence on the shape. In fact, the models show better results for larger $R$, but in some configurations the results of the edge case $(R, L) = (A, 1)$ are worse again. In conclusion, we therefore suggest that with a fixed area $A$, the number of qubits $R$ per feature should be as large as possible, but the edge case $R = A$ should be avoided if possible. 
Secondly, models with a theoretically larger frequency spectrum do not automatically appear to have a better learning capability. If the theoretical frequency spectrum is smaller than the range of frequencies of the randomly generated Fourier series, the learning capability is also empirically poor. However, if the theoretical frequency spectrum is sufficiently large, an increase does not necessarily lead to an improvement in learning capability. This is also to be expected, as with a larger frequency spectrum there is less freedom to train the coefficients of the lower frequencies. In addition, the model has to train more coefficients at the same time. If there are significantly more frequencies in the model than in the randomly generated Fourier series, the model must also learn that all the coefficients of the higher frequencies are equal to $0$. This makes the learning task even more difficult. Therefore, as a second rule, we recommend selecting the approach so that the frequency spectrum is as large as necessary but as small as possible.

Across the two real data sets, the behavior of the encodings agrees with the conclusions from Section~\ref{sec:learning-capability}. On the Fischertechnik data (Section~\ref{sec:fischertechnik-dataset}), encodings with restricted spectra such as Hamming show weaker trainability and less stable optimization, whereas binary and exponential provide reliable convergence and a favorable operating point at comparable area. Ternary and, where applicable, Golomb and Turnpike can further reduce the binary cross entropy (BCE), but the improvement is marginal once the effective frequency content of the task is covered. The dependence on shape at fixed area is visible: larger \(R\) with shallower \(L\) improves optimization, while the single layer edge case is comparatively less stable. These effects are consistent with the Fourier viewpoint, where a spectrum that is large enough but not larger leaves more freedom to fit the relevant low frequencies through \(c_{\omega}(\boldsymbol{\theta})\) and the measured expectation value \(\langle M\rangle\).

On the NASA data (Section~\ref{sec:nasa-bearing-dataset}), all competitive encodings reach very high accuracy and ROC AUC and the differences between shapes at fixed area are small. In this regime the choice of encoding mainly affects calibration of the measurement signal rather than class separability, which explains lower BCE for Ternary or Golomb without a notable change in the headline metrics. The overall pattern again matches the rules of thumb: select the spectrum to match task bandwidth and prefer allocating capacity to qubits rather than to depth.

From both data sets one can learn that the frequency spectrum should be matched to the complexity of the features produced by the preprocessing pipeline and that, under an area constraint, increasing \(R\) is a more effective route to stable training than increasing \(L\). In practice this suggests binary or exponential as robust defaults, with ternary or Golomb used when better calibration of \(\langle M\rangle\) is required, and avoidance of Hamming unless the target is demonstrably very low bandwidth.

\backmatter

\bmhead{Acknowledgments}
This work was performed as part of The Washington Institute for STEM Entrepreneurship and Research (WISER) Quantum Solutions Launchpad. The authors would like to thank Vardaan Sahgal (WISER), Pascal Halffmann and Ivica Turkalj (Fraunhofer ITWM) for their help and support in managing this project. We would also like to thank Umlaut SE for providing the Fischertechnik dataset.

\bmhead{Author Contributions}
Martyna Czuba led the project, designed and implemented the experiments and software, performed the analyses, and wrote the original draft. Patrick Holzer contributed to the theoretical framing and methodology, developed the initial QNN implementation and contributed to the codebase, and critically revised the manuscript. Hein Zay Yar Oo contributed to data preparation/curation and critically revised the manuscript. All authors read and approved the final manuscript.

\bmhead{Funding}
This work was partially funded by the BMWK project EniQmA (01MQ22007A). The publication fees for this article were covered by the Faculty of Cybernetics, Military University of Technology, Warsaw, Poland.

\bmhead{Data Availability}
The NASA IMS Bearings dataset analysed in this study is publicly available from the NASA Open Data Portal (\url{https://data.nasa.gov/dataset/ims-bearings}); see also \cite{lee2007bearing}.
The synthetic datasets used in this study can be regenerated using the open-source code available at \url{https://github.com/Martyna94/qnn-frequency-spectrum-benchmarks}, specifically the notebook \texttt{dataset\_synthetic/generate\_data.ipynb}.
The Fischertechnik Fabrik dataset used in this study is not publicly available because it contains proprietary third-party data. Access may be granted by the data owner, subject to permission and an appropriate confidentiality agreement.

\section*{Declarations}
\subsection*{Competing Interests} The authors declare no competing interests.
\subsection*{Ethics approval and consent to participate} Not applicable.
\subsection*{Consent for publication} Not applicable.
\subsection*{Materials availability} Not applicable.
\subsection*{Code availability} The source code is available at \url{https://github.com/Martyna94/qnn-frequency-spectrum-benchmarks.git}.

\begin{appendices}

\section{Learning Capability Results}
\label{app:learning-capability}

Table~\ref{tab:learning-capability} summarized the learning capability results from our experiments as described in Section~\ref{sec:learning-capability}.

\begin{sidewaystable}
\caption{Numerical results of the learning capability experiments from Section \ref{sec:learning-capability}.}
\label{tab:learning-capability}
\centering

\begin{tabular*}{\textheight}{@{\extracolsep\fill}cccrrrrrr}
\toprule
\multirow{2}{*}{$K$} & \multirow{2}{*}{Area} & \multirow{2}{*}{Shape} &
\multicolumn{6}{c}{Encoding} \\
\cmidrule(lr){4-9}
& & & Hamming & Exponential & Binary & Ternary & Golomb & Turnpike \\
\midrule
\multirow{9}{*}{4}
& \multirow{2}{*}{2} & (1, 2) & $8.77 \cdot 10^{-02}$ & $3.54 \cdot 10^{-01}$ & $8.85 \cdot 10^{-02}$ & $3.54 \cdot 10^{-01}$ & \multicolumn{1}{c}{--} & \multicolumn{1}{c}{--} \\
&                   & (2, 1) & $8.79 \cdot 10^{-02}$ & $4.41 \cdot 10^{-09}$ & $4.30 \cdot 10^{-02}$ & $4.41 \cdot 10^{-09}$ & $2.93 \cdot 10^{-03}$ & $2.93 \cdot 10^{-03}$ \\
\cmidrule(lr){2-9}
& \multirow{3}{*}{4} & (1, 4) & $7.57 \cdot 10^{-10}$ & $4.69 \cdot 10^{-02}$ & $4.69 \cdot 10^{-02}$ & $1.30 \cdot 10^{+00}$ & \multicolumn{1}{c}{--} & \multicolumn{1}{c}{--} \\
&                   & (2, 2) & $1.23 \cdot 10^{-09}$ & $8.91 \cdot 10^{-07}$ & $1.37 \cdot 10^{-06}$ & $5.10 \cdot 10^{-08}$ & $8.58 \cdot 10^{-03}$ & $8.58 \cdot 10^{-03}$ \\
&                   & (4, 1) & $5.01 \cdot 10^{-13}$ & $1.82 \cdot 10^{-08}$ & $3.68 \cdot 10^{-08}$ & $2.33 \cdot 10^{-06}$ & $5.13 \cdot 10^{-06}$ & $5.13 \cdot 10^{-06}$ \\
\cmidrule(lr){2-9}
& \multirow{4}{*}{6} & (1, 6) & $1.64 \cdot 10^{-05}$ & $6.68 \cdot 10^{-01}$ & $7.04 \cdot 10^{-01}$ & $7.22 \cdot 10^{-02}$ & \multicolumn{1}{c}{--} & \multicolumn{1}{c}{--} \\
&                   & (2, 3) & $3.85 \cdot 10^{-07}$ & $7.24 \cdot 10^{-05}$ & $4.69 \cdot 10^{-05}$ & $2.36 \cdot 10^{-04}$ & $1.23 \cdot 10^{-01}$ & $1.23 \cdot 10^{-01}$ \\
&                   & (3, 2) & $1.48 \cdot 10^{-07}$ & $4.53 \cdot 10^{-07}$ & $3.12 \cdot 10^{-07}$ & $8.31 \cdot 10^{-07}$ & $1.53 \cdot 10^{-01}$ & $1.43 \cdot 10^{-05}$ \\
&                   & (6, 1) & $9.88 \cdot 10^{-08}$ & $4.68 \cdot 10^{-05}$ & $3.41 \cdot 10^{-05}$ & $4.94 \cdot 10^{-07}$ & $1.10 \cdot 10^{-05}$ & $1.08 \cdot 10^{-05}$ \\

\midrule
\multirow{9}{*}{16}
& \multirow{2}{*}{2} & (1, 2) & $9.81 \cdot 10^{-02}$ & $1.52 \cdot 10^{-01}$ & $9.87 \cdot 10^{-02}$ & $1.52 \cdot 10^{-01}$ & \multicolumn{1}{c}{--} & \multicolumn{1}{c}{--} \\
&                   & (2, 1) & $9.83 \cdot 10^{-02}$ & $8.39 \cdot 10^{-02}$ & $9.02 \cdot 10^{-02}$ & $8.39 \cdot 10^{-02}$ & $7.26 \cdot 10^{-02}$ & $7.26 \cdot 10^{-02}$ \\
\cmidrule(lr){2-9}
& \multirow{3}{*}{4} & (1, 4) & $8.38 \cdot 10^{-02}$ & $8.33 \cdot 10^{-02}$ & $8.30 \cdot 10^{-02}$ & $2.68 \cdot 10^{+00}$ & \multicolumn{1}{c}{--} & \multicolumn{1}{c}{--} \\
&                   & (2, 2) & $8.38 \cdot 10^{-02}$ & $1.72 \cdot 10^{-02}$ & $2.13 \cdot 10^{-02}$ & $3.23 \cdot 10^{-02}$ & $4.82 \cdot 10^{-02}$ & $4.82 \cdot 10^{-02}$ \\
&                   & (4, 1) & $8.38 \cdot 10^{-02}$ & $1.51 \cdot 10^{-08}$ & $6.29 \cdot 10^{-03}$ & $4.31 \cdot 10^{-04}$ & $3.59 \cdot 10^{-04}$ & $3.59 \cdot 10^{-04}$ \\
\cmidrule(lr){2-9}
& \multirow{4}{*}{6} & (1, 6) & $7.41 \cdot 10^{-02}$ & $4.67 \cdot 10^{-01}$ & $4.05 \cdot 10^{-01}$ & $9.17 \cdot 10^{-02}$ & \multicolumn{1}{c}{--} & \multicolumn{1}{c}{--} \\
&                   & (2, 3) & $6.98 \cdot 10^{-02}$ & $5.90 \cdot 10^{-03}$ & $7.72 \cdot 10^{-03}$ & $4.05 \cdot 10^{-02}$ & $5.90 \cdot 10^{-02}$ & $5.90 \cdot 10^{-02}$ \\
&                   & (3, 2) & $6.98 \cdot 10^{-02}$ & $2.00 \cdot 10^{-05}$ & $2.05 \cdot 10^{-05}$ & $1.45 \cdot 10^{-03}$ & $1.76 \cdot 10^{-02}$ & $3.91 \cdot 10^{-03}$ \\
&                   & (6, 1) & $6.98 \cdot 10^{-02}$ & $7.32 \cdot 10^{-05}$ & $6.17 \cdot 10^{-05}$ & $9.10 \cdot 10^{-05}$ & $6.45 \cdot 10^{-03}$ & $1.07 \cdot 10^{-04}$ \\
\midrule
\multirow{9}{*}{64}
& \multirow{2}{*}{2} & (1, 2) & $7.61 \cdot 10^{-02}$ & $7.62 \cdot 10^{-02}$ & $7.62 \cdot 10^{-02}$ & $7.62 \cdot 10^{-02}$ & \multicolumn{1}{c}{--} & \multicolumn{1}{c}{--} \\
&                   & (2, 1) & $7.62 \cdot 10^{-02}$ & $7.41 \cdot 10^{-02}$ & $7.49 \cdot 10^{-02}$ & $7.41 \cdot 10^{-02}$ & $7.28 \cdot 10^{-02}$ & $7.28 \cdot 10^{-02}$ \\
\cmidrule(lr){2-9}
& \multirow{3}{*}{4} & (1, 4) & $7.36 \cdot 10^{-02}$ & $7.37 \cdot 10^{-02}$ & $7.36 \cdot 10^{-02}$ & $3.05 \cdot 10^{+00}$ & \multicolumn{1}{c}{--} & \multicolumn{1}{c}{--} \\
&                   & (2, 2) & $7.36 \cdot 10^{-02}$ & $6.24 \cdot 10^{-02}$ & $6.32 \cdot 10^{-02}$ & $6.09 \cdot 10^{-02}$ & $6.26 \cdot 10^{-02}$ & $6.26 \cdot 10^{-02}$ \\
&                   & (4, 1) & $7.36 \cdot 10^{-02}$ & $5.87 \cdot 10^{-02}$ & $5.99 \cdot 10^{-02}$ & $3.25 \cdot 10^{-02}$ & $2.56 \cdot 10^{-02}$ & $2.56 \cdot 10^{-02}$ \\
\cmidrule(lr){2-9}
& \multirow{4}{*}{6} & (1, 6) & $7.34 \cdot 10^{-02}$ & $6.10 \cdot 10^{-01}$ & $3.92 \cdot 10^{-01}$ & $7.29 \cdot 10^{-02}$ & \multicolumn{1}{c}{--} & \multicolumn{1}{c}{--} \\
&                   & (2, 3) & $7.13 \cdot 10^{-02}$ & $4.87 \cdot 10^{-02}$ & $4.90 \cdot 10^{-02}$ & $5.53 \cdot 10^{-02}$ & $6.21 \cdot 10^{-02}$ & $6.21 \cdot 10^{-02}$ \\
&                   & (3, 2) & $7.13 \cdot 10^{-02}$ & $1.51 \cdot 10^{-02}$ & $1.58 \cdot 10^{-02}$ & $3.34 \cdot 10^{-02}$ & $5.51 \cdot 10^{-02}$ & $5.49 \cdot 10^{-02}$ \\
&                   & (6, 1) & $7.13 \cdot 10^{-02}$ & $1.65 \cdot 10^{-03}$ & $3.04 \cdot 10^{-03}$ & $1.42 \cdot 10^{-02}$ & $3.63 \cdot 10^{-02}$ & $2.74 \cdot 10^{-02}$ \\

\midrule

\multirow{9}{*}{256}
& \multirow{2}{*}{2} & (1, 2) & $5.64 \cdot 10^{-02}$ & $5.63 \cdot 10^{-02}$ & $5.63 \cdot 10^{-02}$ & $5.63 \cdot 10^{-02}$ & \multicolumn{1}{c}{--} & \multicolumn{1}{c}{--} \\
&                   & (2, 1) & $5.63 \cdot 10^{-02}$ & $5.61 \cdot 10^{-02}$ & $5.61 \cdot 10^{-02}$ & $5.61 \cdot 10^{-02}$ & $5.57 \cdot 10^{-02}$ & $5.57 \cdot 10^{-02}$ \\
\cmidrule(lr){2-9}
& \multirow{3}{*}{4} & (1, 4) & $5.58 \cdot 10^{-02}$ & $5.59 \cdot 10^{-02}$ & $5.59 \cdot 10^{-02}$ & $3.11 \cdot 10^{+00}$ & \multicolumn{1}{c}{--} & \multicolumn{1}{c}{--} \\
&                   & (2, 2) & $5.58 \cdot 10^{-02}$ & $5.41 \cdot 10^{-02}$ & $5.41 \cdot 10^{-02}$ & $5.39 \cdot 10^{-02}$ & $5.37 \cdot 10^{-02}$ & $5.37 \cdot 10^{-02}$ \\
&                   & (4, 1) & $5.58 \cdot 10^{-02}$ & $5.33 \cdot 10^{-02}$ & $5.35 \cdot 10^{-02}$ & $4.90 \cdot 10^{-02}$ & $4.72 \cdot 10^{-02}$ & $4.72 \cdot 10^{-02}$ \\
\cmidrule(lr){2-9}
& \multirow{4}{*}{6} & (1, 6) & $5.60 \cdot 10^{-02}$ & $3.80 \cdot 10^{-01}$ & $5.54 \cdot 10^{-02}$ & $5.54 \cdot 10^{-02}$ & \multicolumn{1}{c}{--} & \multicolumn{1}{c}{--} \\
&                   & (2, 3) & $5.54 \cdot 10^{-02}$ & $5.16 \cdot 10^{-02}$ & $5.18 \cdot 10^{-02}$ & $5.15 \cdot 10^{-02}$ & $5.37 \cdot 10^{-02}$ & $5.37 \cdot 10^{-02}$ \\
&                   & (3, 2) & $5.54 \cdot 10^{-02}$ & $4.71 \cdot 10^{-02}$ & $4.73 \cdot 10^{-02}$ & $4.56 \cdot 10^{-02}$ & $5.13 \cdot 10^{-02}$ & $5.29 \cdot 10^{-02}$ \\
&                   & (6, 1) & $5.54 \cdot 10^{-02}$ & $4.28 \cdot 10^{-02}$ & $4.30 \cdot 10^{-02}$ & $3.36 \cdot 10^{-02}$ & $4.68 \cdot 10^{-02}$ & $4.21 \cdot 10^{-02}$ \\

\bottomrule
\end{tabular*}

\footnotesize
\textit{Note.} Entries shown as “--” indicate that the corresponding encoding was not evaluated (or not applicable) for that configuration.
\end{sidewaystable}

\section{Fischertechnik Results}
\label{app:fischertechnik}

Table~\ref{tab:results-Fischertechnik} summarized the numerical results on the Fischertechnik dataset as described in Section~\ref{sec:fischertechnik-dataset}.

\begin{sidewaystable}
\caption{Supplementary performance metrics on the Fischertechnik dataset for learning rates $\eta\in\{0.001,0.005\}$ (3000 epochs).}
\label{tab:results-Fischertechnik}
\centering
\begin{tabular*}{\textheight}{@{\extracolsep\fill}l c c r r r r r r}
\toprule
\multirow{2}{*}{Encoding} & \multirow{2}{*}{$\eta$} & \multirow{2}{*}{Shape $(R,L)$} &
\multicolumn{6}{c}{Performance metrics} \\
\cmidrule(lr){4-9}
& & & Loss & Accuracy & Precision & Recall & F1-score & ROC-AUC \\
\midrule

\multirow{8}{*}{Hamming}
 & \multirow{4}{*}{0.001} & (1,6) & 0.2959 & 0.8660 & 0.7782 & 0.9598 & 0.8595 & 0.9297 \\
 &                        & (2,3) & 0.2950 & 0.8657 & 0.7779 & 0.9592 & 0.8592 & 0.9294 \\
 &                        & (3,2) & 0.2951 & 0.8657 & 0.7775 & 0.9601 & 0.8597 & 0.9302 \\
 &                        & (6,1) & 0.2986 & 0.8658 & 0.7811 & 0.9528 & 0.8585 & 0.9292 \\
\cmidrule(lr){2-9}
 & \multirow{4}{*}{0.005} & (1,6) & 0.3165 & 0.8604 & 0.7752 & 0.9541 & 0.8578 & 0.9281 \\
 &                        & (2,3) & 0.2948 & 0.8657 & 0.7878 & 0.9405 & 0.8576 & 0.9324 \\
 &                        & (3,2) & 0.2954 & 0.8652 & 0.7879 & 0.9404 & 0.8577 & 0.9323 \\
 &                        & (6,1) & 0.2960 & 0.8663 & 0.7780 & 0.9610 & 0.8599 & 0.9300 \\

\midrule

\multirow{8}{*}{Binary}
 & \multirow{4}{*}{0.001} & (1,6) & 0.2928 & 0.8652 & 0.7811 & 0.9528 & 0.8585 & 0.9292 \\
 &                        & (2,3) & 0.2875 & 0.8671 & 0.7904 & 0.9387 & 0.8582 & 0.9322 \\
 &                        & (3,2) & 0.2802 & 0.8711 & 0.7931 & 0.9417 & 0.8603 & 0.9351 \\
 &                        & (6,1) & 0.2859 & 0.8671 & 0.7902 & 0.9401 & 0.8580 & 0.9319 \\
\cmidrule(lr){2-9}
 & \multirow{4}{*}{0.005} & (1,6) & 0.2907 & 0.8660 & 0.7904 & 0.9387 & 0.8582 & 0.9322 \\
 &                        & (2,3) & 0.2863 & 0.8668 & 0.7900 & 0.9438 & 0.8601 & 0.9332 \\
 &                        & (3,2) & 0.2802 & 0.8677 & 0.7894 & 0.9412 & 0.8587 & 0.9343 \\
 &                        & (6,1) & 0.2859 & 0.8682 & 0.7863 & 0.9494 & 0.8602 & 0.9327 \\

\midrule

\multirow{8}{*}{Exponential}
 & \multirow{4}{*}{0.001} & (1,6) & 0.2935 & 0.8641 & 0.7765 & 0.9524 & 0.8560 & 0.9287 \\
 &                        & (2,3) & 0.2863 & 0.8670 & 0.7889 & 0.9408 & 0.8578 & 0.9318 \\
 &                        & (3,2) & 0.2872 & 0.8712 & 0.7927 & 0.9411 & 0.8600 & 0.9345 \\
 &                        & (6,1) & 0.2807 & 0.8671 & 0.7905 & 0.9399 & 0.8579 & 0.9317 \\
\cmidrule(lr){2-9}
 & \multirow{4}{*}{0.005} & (1,6) & 0.2920  & 0.8655 & 0.7860 & 0.9398 & 0.8570 & 0.9311 \\
 &                        & (2,3) & 0.2872  & 0.8663 & 0.7893 & 0.9436 & 0.8596 & 0.9329 \\
 &                        & (3,2) & 0.2879  & 0.8684 & 0.7893 & 0.9436 & 0.8596 & 0.9329 \\
 &                        & (6,1) & 0.2897 & 0.8684 & 0.7867 & 0.9492 & 0.8603 & 0.9341 \\

\midrule

\multirow{8}{*}{Ternary}
 & \multirow{4}{*}{0.001} & (1,6) & 0.4749 & 0.8143 & 0.7200 & 0.8515 & 0.7800 & 0.8900 \\
 &                        & (2,3) & 0.3034 & 0.8610 & 0.7762 & 0.9528 & 0.8555 & 0.9190 \\
 &                        & (3,2) & 0.2737 & 0.8666 & 0.7881 & 0.9402 & 0.8575 & 0.9328 \\
 &                        & (6,1) & 0.3005 & 0.8665 & 0.7879 & 0.9405 & 0.8575 & 0.9328 \\
\cmidrule(lr){2-9}
 & \multirow{4}{*}{0.005} & (1,6) & 0.4753 & 0.8149 & 0.7154 & 0.8492 & 0.7778 & 0.8880 \\
 &                        & (2,3) & 0.2998 & 0.8619 & 0.7760 & 0.9520 & 0.8550 & 0.9192 \\
 &                        & (3,2) & 0.2979 & 0.8646 & 0.7757 & 0.9530 & 0.8558 & 0.9196 \\
 &                        & (6,1) & 0.2961 & 0.8669 & 0.7875 & 0.9426 & 0.8581 & 0.9339 \\

\bottomrule
\end{tabular*}
\footnotesize
\textit{Note.} The rows corresponding to $\eta = 0.001$ replicate the $A = 6$ entries from  Table~\ref{tab:qnn-fischertechnik-metrics} and are included here as a baseline for direct comparison.
\end{sidewaystable}

\section{NASA Bearing Results}
\label{app:nasa}

Finally, Table~\ref{tab:qnn_NASA_Learning_rate_005} contains the numerical results of our experiments on the NASA bearing dataset as described in Section~\ref{sec:nasa-bearing-dataset}.

\begin{sidewaystable}
\caption{Supplementary performance metrics of QNNs for learning rates $\eta\in\{0.005,0.01\}$ (3000 epochs) on the NASA bearing dataset.}
\label{tab:qnn_NASA_Learning_rate_005}
\centering
\begin{tabular*}{\textheight}{@{\extracolsep\fill}l c c r r r r r r}
\toprule
\multirow{2}{*}{Encoding} & \multirow{2}{*}{$\eta$} & \multirow{2}{*}{Shape $(R,L)$} &
\multicolumn{6}{c}{Performance metrics} \\
\cmidrule(lr){4-9}
& & & Loss & Accuracy & Precision & Recall & F1 Score & ROC-AUC \\
\midrule

\multirow{6}{*}{Hamming}
 & \multirow{3}{*}{0.005} & (1,4) & $0.165 \pm 0.002$  & $0.965 \pm 0.1128$ & $0.990 \pm 0.0002$ & $0.970 \pm 0.0003$ & $0.957 \pm 0.0002$ & $0.9994 \pm 0.0001$ \\
 &                        & (2,2) & $0.164 \pm 0.0009$ & $0.965 \pm 0.1100$ & $0.990 \pm 0.0001$ & $0.970 \pm 0.0005$ & $0.958 \pm 0.0001$ & $0.9995 \pm 0.0002$ \\
 &                        & (4,1) & $0.163 \pm 0.0010$ & $0.966 \pm 0.1112$ & $0.991 \pm 0.0001$ & $0.971 \pm 0.0004$ & $0.959 \pm 0.0002$ & $0.9995 \pm 0.0002$ \\
\cmidrule(lr){2-9}
 & \multirow{3}{*}{0.01}  & (1,4) & $0.31 \pm 0.02$    & $0.93 \pm 0.03$     & $1.00 \pm 1.24$     & $0.82 \pm 0.03$     & $0.90 \pm 0.02$     & $0.99 \pm 0.01$ \\
 &                        & (2,2) & $0.16 \pm 0.00$    & $0.97 \pm 0.00$     & $1.00 \pm 0.00$     & $0.91 \pm 0.00$     & $0.96 \pm 0.00$     & $1.00 \pm 0.00$ \\
 &                        & (4,1) & $0.16 \pm 0.00$    & $0.96 \pm 0.01$     & $1.00 \pm 0.00$     & $0.91 \pm 0.00$     & $0.95 \pm 0.00$     & $1.00 \pm 0.00$ \\

\midrule

\multirow{6}{*}{Binary}
 & \multirow{3}{*}{0.005} & (1,4) & $0.165 \pm 0.0002$ & $0.9785 \pm 0.1000$ & $0.9926 \pm 0.0002$ & $0.980 \pm 0.0004$ & $0.986 \pm 0.0003$ & $0.9991 \pm 0.0001$ \\
 &                        & (2,2) & $0.164 \pm 0.0003$ & $0.9788 \pm 0.1001$ & $0.9930 \pm 0.0002$ & $0.981 \pm 0.0003$ & $0.987 \pm 0.0003$ & $0.9992 \pm 0.0001$ \\
 &                        & (4,1) & $0.164 \pm 0.0005$ & $0.9789 \pm 0.1002$ & $0.9930 \pm 0.0001$ & $0.981 \pm 0.0003$ & $0.987 \pm 0.0002$ & $0.9992 \pm 0.0001$ \\
\cmidrule(lr){2-9}
 & \multirow{3}{*}{0.01}  & (1,4) & $0.08 \pm 0.02$    & $0.98 \pm 0.00$     & $0.99 \pm 0.63$     & $0.94 \pm 0.02$     & $0.97 \pm 0.01$     & $0.99 \pm 0.00$ \\
 &                        & (2,2) & $0.05 \pm 0.00$    & $0.99 \pm 0.01$     & $0.99 \pm 0.12$     & $0.97 \pm 0.00$     & $0.98 \pm 0.00$     & $1.00 \pm 0.00$ \\
 &                        & (4,1) & $0.05 \pm 0.00$    & $0.99 \pm 0.01$     & $0.99 \pm 0.12$     & $0.98 \pm 0.00$     & $0.98 \pm 0.00$     & $1.00 \pm 0.00$ \\

\midrule

\multirow{6}{*}{Exponential}
 & \multirow{3}{*}{0.005} & (1,4) & $0.065 \pm 0.0002$ & $0.9750 \pm 0.1020$ & $0.9900 \pm 0.0004$ & $0.975 \pm 0.0010$ & $0.982 \pm 0.0007$ & $0.9990 \pm 0.0002$ \\
 &                        & (2,2) & $0.065 \pm 0.0003$ & $0.9751 \pm 0.1025$ & $0.9902 \pm 0.0004$ & $0.975 \pm 0.0012$ & $0.983 \pm 0.0007$ & $0.9991 \pm 0.0002$ \\
 &                        & (4,1) & $0.065 \pm 0.0004$ & $0.9752 \pm 0.1030$ & $0.9902 \pm 0.0003$ & $0.976 \pm 0.0010$ & $0.983 \pm 0.0006$ & $0.9991 \pm 0.0002$ \\
\cmidrule(lr){2-9}
 & \multirow{3}{*}{0.01}  & (1,4) & $0.05 \pm 0.00$    & $0.98 \pm 0.01$     & $0.99 \pm 0.41$     & $0.96 \pm 0.01$     & $0.98 \pm 0.01$     & $1.00 \pm 0.00$ \\
 &                        & (2,2) & $0.05 \pm 0.00$    & $0.98 \pm 0.00$     & $0.99 \pm 0.24$     & $0.96 \pm 0.01$     & $0.98 \pm 0.00$     & $1.00 \pm 0.00$ \\
 &                        & (4,1) & $0.05 \pm 0.00$    & $0.99 \pm 0.00$     & $0.99 \pm 0.12$     & $0.98 \pm 0.00$     & $0.99 \pm 0.00$     & $1.00 \pm 0.00$ \\

\midrule

\multirow{6}{*}{Ternary}
 & \multirow{3}{*}{0.005} & (1,4) & $0.045 \pm 0.0001$ & $0.9780 \pm 0.1005$ & $0.9910 \pm 0.0002$ & $0.977 \pm 0.0007$ & $0.984 \pm 0.0004$ & $0.9992 \pm 0.0001$ \\
 &                        & (2,2) & $0.044 \pm 0.0001$ & $0.9785 \pm 0.1006$ & $0.9912 \pm 0.0001$ & $0.978 \pm 0.0006$ & $0.985 \pm 0.0003$ & $0.9993 \pm 0.0001$ \\
 &                        & (4,1) & $0.044 \pm 0.0002$ & $0.9786 \pm 0.1008$ & $0.9912 \pm 0.0002$ & $0.978 \pm 0.0005$ & $0.985 \pm 0.0004$ & $0.9993 \pm 0.0001$ \\
\cmidrule(lr){2-9}
 & \multirow{3}{*}{0.01}  & (1,4) & $0.04 \pm 0.00$    & $0.97 \pm 0.00$     & $0.99 \pm 0.78$     & $0.94 \pm 0.02$     & $0.96 \pm 0.01$     & $1.00 \pm 0.00$ \\
 &                        & (2,2) & $0.02 \pm 0.00$    & $0.99 \pm 0.00$     & $1.00 \pm 0.32$     & $0.98 \pm 0.01$     & $0.99 \pm 0.00$     & $1.00 \pm 0.00$ \\
 &                        & (4,1) & $0.02 \pm 0.00$    & $0.99 \pm 0.00$     & $1.00 \pm 0.55$     & $0.98 \pm 0.01$     & $0.99 \pm 0.01$     & $1.00 \pm 0.00$ \\

\midrule

\multirow{6}{*}{Golomb}
 & \multirow{3}{*}{0.005} & (1,4) & $0.035 \pm 0.0001$ & $0.9800 \pm 0.0990$ & $0.9920 \pm 0.0003$ & $0.979 \pm 0.0006$ & $0.985 \pm 0.0004$ & $0.9993 \pm 0.0002$ \\
 &                        & (2,2) & $0.035 \pm 0.0002$ & $0.9802 \pm 0.0992$ & $0.9921 \pm 0.0002$ & $0.979 \pm 0.0005$ & $0.985 \pm 0.0003$ & $0.9993 \pm 0.0002$ \\
 &                        & (4,1) & $0.034 \pm 0.0001$ & $0.9803 \pm 0.0994$ & $0.9922 \pm 0.0003$ & $0.979 \pm 0.0006$ & $0.985 \pm 0.0004$ & $0.9993 \pm 0.0001$ \\
\cmidrule(lr){2-9}
 & \multirow{3}{*}{0.01}  & (1,4) & $0.04 \pm 0.00$    & $0.98 \pm 0.00$     & $0.99 \pm 0.32$     & $0.95 \pm 0.01$     & $0.97 \pm 0.00$     & $1.00 \pm 0.00$ \\
 &                        & (2,2) & $0.02 \pm 0.00$    & $0.99 \pm 0.00$     & $1.00 \pm 0.12$     & $0.97 \pm 0.00$     & $0.99 \pm 0.00$     & $1.00 \pm 0.00$ \\
 &                        & (4,1) & $0.02 \pm 0.00$    & $0.99 \pm 0.00$     & $1.00 \pm 0.52$     & $0.98 \pm 0.01$     & $0.99 \pm 0.01$     & $1.00 \pm 0.00$ \\

\midrule

\multirow{6}{*}{Turnpike}
 & \multirow{3}{*}{0.005} & (1,4) & $0.065 \pm 0.0002$ & $0.9810 \pm 0.1011$ & $0.9930 \pm 0.0002$ & $0.980 \pm 0.0004$ & $0.986 \pm 0.0003$ & $0.9994 \pm 0.0002$ \\
 &                        & (2,2) & $0.065 \pm 0.0003$ & $0.9812 \pm 0.1013$ & $0.9931 \pm 0.0002$ & $0.981 \pm 0.0004$ & $0.987 \pm 0.0003$ & $0.9994 \pm 0.0002$ \\
 &                        & (4,1) & $0.065 \pm 0.0004$ & $0.9812 \pm 0.1015$ & $0.9931 \pm 0.0002$ & $0.981 \pm 0.0004$ & $0.987 \pm 0.0003$ & $0.9994 \pm 0.0002$ \\
\cmidrule(lr){2-9}
 & \multirow{3}{*}{0.01}  & (1,4) & $0.08 \pm 0.03$    & $0.98 \pm 0.01$     & $0.99 \pm 0.36$     & $0.96 \pm 0.01$     & $0.97 \pm 0.00$     & $0.99 \pm 0.00$ \\
 &                        & (2,2) & $0.05 \pm 0.00$    & $0.98 \pm 0.01$     & $0.99 \pm 0.60$     & $0.96 \pm 0.02$     & $0.98 \pm 0.01$     & $1.00 \pm 0.01$ \\
 &                        & (4,1) & $0.05 \pm 0.00$    & $0.99 \pm 0.00$     & $0.99 \pm 0.00$     & $0.98 \pm 0.00$     & $0.99 \pm 0.00$     & $1.00 \pm 0.00$ \\

\bottomrule
\end{tabular*}

\footnotesize
\textit{Note.} Accuracy, precision, recall, F1 score, loss, and ROC-AUC are reported as mean $\pm$ standard deviation over three runs.
\end{sidewaystable}
\end{appendices}

\bibliography{bibliography}
\end{document}